\documentclass{article}%
\usepackage{amsmath}
\usepackage{amsfonts}
\usepackage{amssymb}
\usepackage{graphicx}%
\newcommand{\bpartial}{\mathop{\partial\kern -4pt\raisebox{.8pt}{$|$}}}
\newcommand{\bra}{\mathopen{[\kern-1.6pt[}}
\newcommand{\ket}{\mathclose{]\kern-1.5pt]}}
\newcommand{\bbra}{\mathopen{[\kern-2.2pt[\kern-2.3pt[}}
\newcommand{\bket}{\mathclose{]\kern-2.1pt]\kern-2.3pt]}}

\makeindex
\begin{document}

\title { Classification of four and six dimensional  Drinfel'd superdoubles}

\vspace{10mm}

\author {  A. Eghbali, \hspace{-3mm}{ \footnote{ e-mail: a.eghbali@azaruniv.edu}}\hspace{2mm}
A. Rezaei-Aghdam \hspace{-2mm}{ \footnote{Corresponding author.
e-mail:
rezaei-a@azaruniv.edu}\hspace{2mm} {\small and}\hspace{2mm} F. Heidarpour} \\
{\small{\em Department of Physics, Faculty of science, Azarbaijan University }}\\
{\small{\em of Tarbiat Moallem , 53714-161, Tabriz, Iran  }}}

\maketitle

\vspace{10mm}

\begin{abstract}
Using adjoint representation we firstly classify  two and three
dimensional Lie super-bialgebras  obtain from decomposable Lie
superalgebras. In this way  we complete the classification
obtained by Eghbali {\it et al.}, [J. Math. Phys. {\bf 51},
073503 (2010); e-print arXiv:0901.4471 [math-ph]]. Then we
classify all four and six dimensional Drinfel'd superdoubles.
\end{abstract}

\newpage
%%%%%%%%%%%%%%%%%%%%%%%%%%%%%%%%%%%%%%%%%%%%%%%%%%%%%%%%%%%%%%%%%%%%%%%%%%%%%%%%%%%%%%%%%%%%%%%55
\section {\large {\bf Introduction}}

Lie super-bialgebras \cite{N.A}, as the underlying symmetry
algebras, play an important role in the integrable structure of
$AdS/CFT$ correspondence \cite{Bs}. Similarly, one can consider
Poisson-Lie T-dual sigma models on Poisson-Lie supergroups
\cite{ER}. In this way and by considering that there is a
universal quantization for Lie super-bialgebras \cite{Geer}, one
can assign an important role to the classification of Lie
super-bialgebras (especially low dimensional Lie super-bialgebras)
from both physical and mathematical point of view. Until now there
are distinguished and nonsystematic ways for obtaining low
dimensional Lie super-bialgebras (see, for example,
\cite{{J.z},{J},{KARAALI}}). In  Ref. \cite{ER1}, using the
adjoint representation of Lie superalgebras, we have  given a
systematic way for obtaining and classification of low
dimensional Lie super-bialgebras and applied this method to the
classification of two and three dimensional Lie super-bialgebras
related to indecomposable Lie superalgebras of Ref. \cite{B}. In
the continuation of that work we tried to classify all Lie
superalgebras of Drinfel'd superdoubles of these Lie
super-bialgebras. Then, we saw Ref. \cite{H} in which they have
tried  to do this work for Lie super-bialgebras related to two
and three dimensional decomposable and indecomposable Lie
superalgebras \cite{B}. But unfortunately there are some
incorrect  results in their paper. Firstly,  because they use same
notation for nonisomorphic Lie superalgebras and incorrect
automorphism Lie supergroups for some Lie superalgebras so the
number of their Lie super-bialgebras are incorrect. Secondly, and
importantly they use nonstandard basis in determination of Lie
superalgebras of Drinfel'd superdoubles. For these reasons some
of their principal results (theorem 3) are incorrect. For
obtaining correct results one must use standard basis for Lie
superalgebras and superdeterminant for isomorphism matrices.
Thirdly, their mixed commutation relations  are not compatible
with ad-invariant form $<.~,~.>$ on $\cal {D}$. For these reasons,
we are motivated to write this paper to complete the results of
Ref. \cite{ER1} and classify two and three dimensional Lie
super-bialgebras related to decomposable Lie superalgebras.
Furthermore  by use of standard basis for Lie superalgebras of
Drinfel'd superdoubles we classify all four and six dimensional
Drinfel'd superdoubles as theorems 1-3.

\smallskip
\goodbreak

The paper is organized as follows. In  section two, we review and
rewrite correctly  the results of Ref. \cite{ER1} about two and
three dimensional Lie super-bialgebras; then by obtaining Lie
super-bialgebras for decomposable Lie superalgebras we complete
that work. In section three, we find nonisomorphic four and six
dimensional Drinfel'd superdoubles by use of standard basis,  the
results are summarized as three theorems. In appendix A we give
some notations about supermatrices and supertensors in the
standard basis. In appendix B we give solutions of super Jacobi
and mixed super Jacobi identities for dual Lie superalgebras of
decomposable Lie superalgebras.  The differences between the list
of Lie super-bialgebras of Ref. \cite{ER1} with those of Ref.
\cite{H} are given in appendix C. In appendix D we give the
(anti)commutation relations of four and six dimensional Drinfel'd
superdoubles. The details of differences between theorem 3 and
those of Ref. \cite{H} are given in appendix E. Finally, the
isomorphism matrices of four and six dimensional Drinfel'd
superdoubles are listed in appendix F.

\vspace{9mm}
%%%%%%%%%%%%%%%%%%%%%%%%%%%%%%%%%%%%%%%%%%%%%%%%%%%%%%%%%%%%%%%%%%%%%%%%%%%%%%%%%%%%%
\section {\large {\bf Two and three dimensional
 Lie super-bialgebras }}

Let us first review some basic definitions and notations about
Lie superalgebras and Lie super-bialgebras (see, Refs. \cite{N.A}
and \cite{ER1}).

\vspace{6mm}

{\it Definition 1}: A {\em Lie superalgebra} ${\bf g}$ is a graded
vector space ${\bf g}={\bf g}_B \oplus {\bf g}_F$ with gradings;
$grade({\bf g}_B)=0,\; grade({\bf g}_F)=1$; such that Lie bracket
satisfies the super antisymmetric and super Jacobi identities,
i.e., in a graded basis $\{X_i\}$ of ${\bf g}$ if we put \cite{R1}

\begin{equation}
[X_i , X_j] = {f^k}_{ij} X_k,
\end{equation}
then
\begin{equation}
(-1)^{i(j+k)}{f^m}_{jl}{f^l}_{ki} + {f^m}_{il}{f^l}_{jk} +
(-1)^{k(i+j)}{f^m}_{kl}{f^l}_{ij}=0,
\end{equation}
so that
\begin{equation}
{f^k}_{ij}=-(-1)^{ij}{f^k}_{ji}.
\end{equation}
Note that, in the conventional basis, ${f^B}_{BB}$ and
${f^F}_{BF}$ are real c-numbers and ${f^B}_{FF}$ are pure
imaginary c-numbers and other components of structure constants
${f^i}_{jk}$ are zero \cite{D}, i.e., we have
\begin{equation}
{f^k}_{ij}=0, \hspace{10mm} if \hspace{5mm} grade(i) +
grade(j)\neq grade(k)\hspace{3mm} (mod 2).
\end{equation}
Let ${\bf g}$ be a finite-dimensional Lie superalgebra and ${\bf
g}^\ast$ be its dual vector space and let $( .~ ,~. )$ be the
canonical pairing on ${\bf g}^\ast \oplus {\bf g}$.

\vspace{6mm}

{\it Definition 2}: A {\em Lie super-bialgebra} structure on a Lie
superalgebra ${\bf g}$ is a super skew-symmetric linear map
$\delta : {\bf g }\longrightarrow {\bf g}\otimes{\bf g}$ (the {\em
super cocommutator}) so that \cite{N.A}

1)~~$\delta$ is a super one-cocycle, i.e.
\begin{equation}
\delta([X,Y])=(ad_X\otimes I+I\otimes
ad_X)\delta(Y)-(-1)^{|X||Y|}(ad_Y\otimes I+I\otimes
ad_Y)\delta(X) \qquad \forall X,Y\in {\bf g},
\end{equation}
~~~~~~~where $|X|(|Y|)$ indicates the grading of $X(Y)$;

2) the dual map ${^t}{\delta}:{\bf g}^\ast\otimes {\bf g}^\ast \to
{\bf g}^\ast$ is a Lie superbracket on ${\bf g}^\ast$, i.e.,
\begin{equation}
(\xi\otimes\eta , \delta(X)) = ({^t}{\delta}(\xi\otimes\eta) , X)
= ([\xi,\eta]_\ast , X)    \qquad \forall X\in  {\bf g} ;\;\,
\xi,\eta\in{\bf g}^\ast.
\end{equation}
The Lie super-bialgebra defined in this way will be indicated by
$({\bf g},{\bf g}^\ast)$ or $({\bf g},\delta)$.

\vspace{6mm}

{\it Proposition 1}: Let $({\bf g} , \delta )$ be a Lie
super-bialgebra. There exists a unique Lie superalgebra
structures with the following commutation relations on the vector
space ${\bf g}\oplus{\bf \tilde{g}}$ such that $\bf g$ and ${\bf
g}^\ast$ are Lie superalgebras and the natural scalar product on
${\bf g}\oplus{\bf \tilde{g}}$ is invariant
\begin{equation}
[x , y]_{\cal{D}}\;=\;[x ,y],\;\; \;[x ,
\xi]_{\cal{D}}\;=\;-(-1)^{|x||\xi|}
ad^{\ast}_{\hspace{0.5mm}\xi}x+ad^{\ast}_{\hspace{0.5mm}x}\xi,\;\;\;[\xi
, \eta]_{\cal{D}}\;=\;[\xi ,\eta]_{{\bf g}^\ast}\;\;\;\;\forall
x,y\in {\bf g};\;\;  \xi, \eta \in {\bf g}^\ast,
\end{equation}
where
\begin{equation}
<ad_x y\;,\;\xi>\;=\;-(-1)^{|x||y|}<y\;,\;
ad^{\ast}_{\hspace{0.5mm}x}\xi>,
\end{equation}
\begin{equation}
<ad_{\xi} \eta \;,\;x>\;=\;-(-1)^{|\xi||\eta|}<\eta\;,\;
ad^{\ast}_{\hspace{0.5mm}\xi}x>.
\end{equation}
The Lie superalgebra  ${\cal{D}} = {\bf g}\oplus{\bf \tilde{g}}$
is called {\em Drinfel'd superdouble}.

\vspace{6mm}

{\it Proposition 2}: If there exists an automorphism $A$ of ${\bf
g}$ such that
\begin{equation}
\delta^\prime = (A\otimes A)\circ\delta\circ A^{-1},
\end{equation}
then the super one-cocycles $\delta$ and $\delta^\prime$ of the
Lie superalgebra $\bf g$ are {\em equivalent}. In this case the
two Lie super-bialgebras $({\bf g},\delta)$ and $({\bf
g},\delta^\prime)$ are equivalent (as in the bosonic case
\cite{RHR}).

\vspace{6mm}

{\it Definition 3}: A {\em Manin super triple} \cite{N.A} is a
triple of Lie superalgebras $(\cal{D} , {\bf g} , {\bf
\tilde{g}})$ together with a non-degenerate ad-invariant super
symmetric bilinear form $<.~,~. >$ on $\cal{D}$, such
that\hspace{2mm}

1)~~${\bf g}$ and ${\bf \tilde{g}}$ are Lie sub-superalgebras of
$\cal{D}$,\hspace{2mm}

2)~ ${\cal{D}} = {\bf g}\oplus{\bf \tilde{g}}$ as a supervector
space,\hspace{2mm}

3)~ ${\bf g}$ and ${\bf \tilde{g}}$ are isotropic with respect to
$<.~,~. >$, i.e.,
\begin{equation}
<X_i , X_j>\; =\; <\tilde{X}^i , \tilde{X}^j> \;=\; 0,
\hspace{10mm} {\delta_i}^j\;=\;<X_i , \tilde{X}^j>\; =\;
(-1)^{ij}<\tilde{X}^j, X_i>\;=\;(-1)^{ij}{\delta^j}_i,
\end{equation}
where $\{X_i\}$ and $\{\tilde{X}^i\}$ are  basis of Lie
superalgebras ${\bf g}$ and ${\bf \tilde{g}}$, respectively. Note
that in the above relation ${\delta^j}_i$ is the ordinary delta
function.  There is a one-to-one correspondence between Lie
super-bialgebra $({\bf g},{\bf g}^\ast)$ and Manin super triple
$(\cal{D} , {\bf g} , {\bf \tilde{g}})$ with $ {\bf g}^\ast={\bf
\tilde{g}}$ \cite{N.A}. If we choose the structure constants of
Lie superalgebras ${\bf g}$ and ${\bf \tilde{g}}$ as
\begin{equation}
[X_i , X_j] = {f^k}_{ij} X_k,\hspace{20mm} [\tilde{X}^i ,\tilde{ X}^j] ={{\tilde{f}}^{ij}}_{\; \; \: k} {\tilde{X}^k}, \\
\end{equation}
then ad-invariance of the bilinear form $<.~,~.>$ on $\cal{D} =
{\bf g}\oplus{\bf \tilde{g}}$ implies that
\begin{equation}
[X_i , \tilde{X}^j] =(-1)^j{\tilde{f}^{jk}}_{\; \; \; \:i} X_k
+(-1)^i {f^j}_{ki} \tilde{X}^k.
\end{equation}
Clearly, using the Eqs. (6), (11) and (12) we have
\begin{equation}
\delta(X_i) = (-1)^{jk}{\tilde{f}^{jk}}_{\; \; \; \:i} X_j \otimes
X_k.
\end{equation}
note that the appearance of $(-1)^{jk}$ in this relation is due
to the definition of natural inner product between ${\bf
g}\otimes {\bf g}$ and  ${\bf g}^\ast \otimes {\bf g}^\ast $ as $
(\tilde{X}^i \otimes \tilde{ X}^j ,X_k \otimes
X_l)=(-1)^{jk}{\delta^i}_k {\delta^j}_l$.\\
As a result, if we apply this relation in the super one-cocycle
condition (5), super Jacobi identities (2) for the  dual Lie
superalgabra and the following mixed super Jacobi identities are
obtained:
\begin{equation}
{f^m}_{jk}{\tilde{f}^{il}}_{\; \; \; \; m}=
{f^i}_{mk}{\tilde{f}^{ml}}_{\; \; \; \; \; j} +
{f^l}_{jm}{\tilde{f}^{im}}_{\; \; \; \; \; k}+ (-1)^{jl}
{f^i}_{jm}{\tilde{f}^{ml}}_{\; \; \; \; \; k}+ (-1)^{ik}
{f^l}_{mk}{\tilde{f}^{im}}_{\; \; \; \; \; j}.
\end{equation}
This relation can also be obtained from super Jacobi identity of
$\cal{D}$. In Ref. \cite{ER1} we find and classify all two and
three dimensional Lie super-bialgebras for all two and three
dimensional indecomposable Lie superalgebras of Ref. \cite{B}. The
method of classification is new and indeed it is improvement and
generalization of the method of Ref. \cite{JR} to the Lie
superalgebras. Note that in Ref. \cite{Snobl} unfortunately there
is not standard and logically method for obtaining of low
dimensional Lie bialgebras. In this method by use of adjoint
representation for super Jacobi and mixed super Jacobi identities
(2) and (15) we write these relations in the matrix form and
solve them for finding dual Lie superalgebras by direct
calculation; then by use of automorphism Lie supergroups of Lie
superalgebras we classify all non isomorphic two and three
dimensional Lie super-bialgebras \cite{ER1}. Here the list of two
and three dimensional indecomposable Lie superalgebras  \cite{B}
and their related Lie super-bialgebras \cite{ER1} are given in
table 1 and tables 4, 5 and 6, respectively; note that as we use
DeWitt notation and standard basis here, the structure constants
$C^B_{FF}$ must be pure imaginary. Furthermore in this section we
consider decomposable Lie superalgebras as in the following table
2 and obtain related Lie super-bialgebras by use of the method
mentioned in Ref. \cite{ER1}.

\vspace{4mm}
%%%%%%%%%%%%%%%%%%%%%%%%%%%%%%%%%%%%%%%%%%%%%%%%%%%%%%%%%%%%%%%%%%%%%%%%%%%%%%%%%5
\begin{center}
\hspace{10mm}{\small {\bf Table 1}} : \hspace{1mm}{ Two and three
dimensional indecomposable  Lie superalgebras (Ref. \cite{R2}).}\\
    \begin{tabular}{l l l l  l l p{25mm} }
    \hline\hline
   \vspace{-1mm}
{\footnotesize Type }& {\footnotesize ${\bf g}$ }& {\footnotesize
Bosonic } & {\footnotesize Fermionic }&{\footnotesize Non-zero
(anti)commutation }&{\footnotesize Comments} \\

& & {\footnotesize  basis} & {\footnotesize basis}&{\footnotesize
relations}& \\\hline

{\footnotesize $(1,1)$}&{\footnotesize $B$}& {\footnotesize $X_1$}& {\footnotesize $X_2$}&
{\footnotesize $[X_1,X_2]=X_2$}&{\footnotesize Trivial} \\

\vspace{1mm}

& {\footnotesize $(A_{1,1}+A)$}& {\footnotesize $X_1$}&
{\footnotesize $X_2$} &{\footnotesize
$\{X_2,X_2\}=iX_1$}&{\footnotesize Nontrivial}
\\

\vspace{1mm}

{\footnotesize $(2,1)$} & {\footnotesize ${C^1_p}$} &
{\footnotesize $X_1,X_2$} & {\footnotesize  $X_3$}&{\footnotesize
$[X_1,X_2]=X_2,\; [X_1,X_3]=pX_3 $}&{\footnotesize   $p\neq0$, Trivial } \\

\vspace{1mm}

&{\footnotesize  $C^1_{\frac{1}{2}}$ } & {\footnotesize $X_1,X_2$}
& {\footnotesize  $X_3$}&{\footnotesize $[X_1,X_2]=X_2,\;
[X_1,X_3]=\frac{1}{2}X_3,\;\{X_3,X_3\}=iX_2 $}&{\footnotesize
Nontrivial}
\\

\vspace{1mm}

{\footnotesize  $(1,2)$}&{\footnotesize   $ C^2_p$} &
{\footnotesize $X_1$} & {\footnotesize  $X_2,X_3$}& {\footnotesize
$[X_1,X_2]=X_2, \;[X_1,X_3]=pX_3$ }&{\footnotesize  $0<|p|\leq 1$, Trivial } \\

\vspace{1mm}

&{\footnotesize $C^3$} & {\footnotesize  $X_1$} & {\footnotesize
$X_2,X_3$}  &{\footnotesize  $[X_1,X_3]=X_2
$}&{\footnotesize Nilpotent, Trivial  } \\

\vspace{1mm}

 &{\footnotesize  $C^4$} & {\footnotesize
$X_1$} & {\footnotesize  $X_2,X_3$}&{\footnotesize  $[X_1,X_2]=X_2,\; [X_1,X_3]=X_2+X_3$}
&{\footnotesize   Trivial }\\

\vspace{1mm}

&{\footnotesize $C^5_p$} & {\footnotesize  $X_1$} & {\footnotesize
$X_2,X_3$}&{\footnotesize $[X_1,X_2]=pX_2-X_3,\;
[X_1,X_3]=X_2+pX_3$ }&{\footnotesize   $p\geq0$, Trivial }
\\

\vspace{1mm}

&{\footnotesize $(A_{1,1}+2A)^1$}& {\footnotesize $X_1$} &
{\footnotesize $X_2,X_3$}&{\footnotesize $\{X_2,X_2\}=iX_1,\;
\{X_3,X_3\}=iX_1 $} &
{\footnotesize Nilpotent, Nontrivial} \\

\vspace{-1mm}

& {\footnotesize $(A_{1,1}+2A)^2$}& {\footnotesize $X_1$} &
{\footnotesize  $X_2,X_3$}& {\footnotesize  $\{X_2,X_2\}=iX_1,
\;\{X_3,X_3\}=-iX_1 $}&{\footnotesize  Nilpotent, Nontrivial }
\smallskip \\
\hline\hline
    \end{tabular}
\end{center}

\vspace{6mm}
%%%%%%%%%%%%%%%%%%%%%%%%%%%%%%%%%%%%%%%%%%%%%%%%%%%%%%%%%%%%%%%%%%%%%%%%%%%%%%%%%%%%%%%555
\begin{center}
\hspace{10mm}{\small {\bf Table 2}} : \hspace{1mm}{  Three
dimensional
decomposable  Lie superalgebras.}\\
    \begin{tabular}{l l l l  l l p{15mm} }
    \hline\hline

\vspace{-1mm}

{\footnotesize Type }& {\footnotesize ${\bf g}$ }& {\footnotesize
Bosonic } & {\footnotesize Fermionic }&{\footnotesize Non-zero
(anti)
commutation }&{\footnotesize Comments} \\

& & {\footnotesize   basis} & {\footnotesize basis}&{\footnotesize
relations}& \\\hline
\vspace{1mm}

{\footnotesize  $(2,1)$} & {\footnotesize  $(B+A_{1,1})$}&
{\footnotesize  $X_1, X_2$}& {\footnotesize  $X_3$}&
{\footnotesize  $[X_1,X_3]=X_3$}& {\footnotesize Solvable, Trivial}\\

\vspace{1mm}

&{\footnotesize  $(2A_{1,1}+A)=(A_{1,1}+A) \oplus A_{1,0}$} &
{\footnotesize $X_1, X_2$} & {\footnotesize $X_3$}&{\footnotesize
$\{X_3,X_3\}=iX_1 $}&{\footnotesize   Nilpotent, Nontrivial}\\

\vspace{1mm}

&{\footnotesize  $C^1_0=C^1_{p=0}$}& {\footnotesize $X_1, X_2$ }&
{\footnotesize $X_3$}&
{\footnotesize $[X_1,X_2]=X_2$}& {\footnotesize Solvable,  Trivial}\\

\vspace{1mm}

{\footnotesize $(1,2)$} & {\footnotesize $C^2_0=C^2_{p=0}=B \oplus
A_{0,1}$} & {\footnotesize $X_1$} & {\footnotesize $ X_2,
X_3$}&{\footnotesize
$[X_1,X_2]=X_2$}& {\scriptsize  Solvable, Trivial}\\

\vspace{-1mm}

&{\footnotesize $(A_{1,1}+2A)^0=(A_{1,1}+A) \oplus A_{0,1}$} &
{\footnotesize $X_1$} & {\footnotesize $ X_2, X_3$}&{\footnotesize
$\{X_2,X_2\}=iX_1$}& {\footnotesize  Nilpotent, Nontrivial}
\smallskip \\
\hline\hline
    \end{tabular}
\end{center}
%%%%%%%%%%%%%%%%%%%%%%%%%%%%%%%%%%%%%%%%%%%%%%%%%%%%%%%%%%%%%%%%%%%%%%%%%%%%%55
\newpage

Note that as  discussed above, for this purpose  we need
automorphism Lie supergroups of these Lie superalgebras where we
have obtained and listed in the following table 3.

\vspace{7mm}

\begin{center}

\hspace{1mm}{\small {\bf Table 3}}: \hspace{1mm}{ Automorphism Lie
supergroups of the two and three dimensional
\\\vspace{-1mm} decomposable Lie superalgebras.
}\\
\begin{tabular}{ l  l  l  p{50cm} }
    \hline\hline
{\footnotesize $\bf g$} & \hspace{2cm}{\footnotesize Automorphism
Lie supergroups}&{\footnotesize Comments}\\ \hline \vspace{2mm}

{\footnotesize $(B+A_{1,1})$} &\hspace{2cm} {\footnotesize $\left(
\begin{array}{ccc}
                      1 & a & 0 \\
                       0 & b & 0 \\
                       0 & 0 & c
                     \end{array} \right)$}& {\footnotesize $b,c\in\Re-\{0\}$ , $a\in\Re$ } \\

\vspace{2mm}

{\footnotesize $(2A_{1,1}+A) $} &\hspace{2cm}{\footnotesize
$\left(\begin{array}{ccc}
                      a^2 & 0 & 0 \\
                       c & b & 0 \\
                       0 & 0 & a
                     \end{array} \right)$}&{\footnotesize $a, b\in\Re-\{0\}$ , $c\in\Re$ }\\

\vspace{2mm}

{\footnotesize $ C^1_0$ }  & \hspace{2cm}{\footnotesize
$\left(\begin{array}{ccc}
                      1 & a & 0 \\
                       0 & b & 0 \\
                       0 & 0 & c
                     \end{array} \right)$}&{\footnotesize $b,c\in\Re-\{0\}$ , $a\in\Re$ }  \\

\vspace{2mm}

{\footnotesize $ C^2_0$}    &\hspace{2cm}{\footnotesize $\left(
\begin{array}{ccc}
                      1 & 0 & 0 \\
                       0 & a & 0 \\
                       0 & 0 & b
                     \end{array} \right)$} &{\footnotesize  $a,b\in\Re-\{0\}$} \\

\vspace{1mm}

{\footnotesize $ (A_{1,1}+2A)^0$}    & \hspace{2cm}{\footnotesize
$\left(
\begin{array}{ccc}
                      a^2 & 0 & 0 \\
                       0 & a & b \\
                       0 &0 & c
                     \end{array} \right)$} & {\footnotesize $a, c\in\Re-\{0\}$, $b\in\Re$}\\\hline\hline
    \end{tabular}
\end{center}

%%%%%%%%%%%%%%%%%%%%%%%%%%%%%%%%%%%%%%%%%%%%%%%%%%%%%%%%%%%%%%%%%%%%%%%%%%%%%%%%%%%%%%%%%%%%%%%%%5
\smallskip

The solutions of super Jacobi and mixed super Jacobi identities
and  isomorphism matrices $C$ related to these decomposable Lie
superalgebras are listed in appendix B. The related Lie
super-bialgebras are listed in tables 4 and 6. Note that in tables
4-6, $I_{(m,n)}$ represent the Abelian Lie superalgebras with
$m(n)$ bosonic(fermionic) generators.

%%%%%%%%%%%%%%%%%%%%%%%%%%%%%%%%%%%%%%%%%%%%%%%%%%%%%%%%%%%%%%%%%%%%%%%%%%%%%
\begin{center}
\hspace{3mm}{\small {\bf Table 4}}: \hspace{1mm}{\small
Three dimensional  Lie super-bialgebras of the type (2 ,1).}\\
    \begin{tabular}{l l l l  p{0.15mm} }
    \hline\hline
{\footnotesize ${\bf g}$ }& {\footnotesize $\tilde{\bf g}$}
&{\footnotesize Non-zero (anti) commutation relations of
$\tilde{\bf g}$}& {\footnotesize Comments} \\ \hline

\vspace{2mm}

{\footnotesize $(2A_{1,1}+A) $}&{\footnotesize $I_{(2 , 1)} $}&
\\

\vspace{1mm}

{\footnotesize $(B+A_{1,1})$}&{\footnotesize$I_{(2 , 1)} $}& &\\

\vspace{1mm}

&{\footnotesize $(B+A_{1,1}).i$}&{\footnotesize $[{\tilde X}^2,{\tilde X}^3]={\tilde X}^3$} & &\\

\vspace{1mm}

 &{\footnotesize $(2A_{1,1}+A)$} &{\footnotesize $\{{\tilde X}^3,{\tilde X}^3\}=i{\tilde X}^1$}
 &\\
\vspace{2mm}

 &{\footnotesize $(2A_{1,1}+A).i$} &{\footnotesize $\{{\tilde X}^3,{\tilde
X}^3\}=-i{\tilde X}^1$} &\\

\vspace{1mm}

{\footnotesize $C^1_p$}&{\footnotesize $I_{(2 , 1)} $}& &{\footnotesize $p\in\Re$ }&\\

\vspace{1mm}

&{\footnotesize $(2A_{1,1}+A)$}&{\footnotesize $\{{\tilde X}^3,{\tilde X}^3\}=i{\tilde X}^1$} &{\footnotesize $p\in\Re$ }\\

\vspace{1mm}

&{\footnotesize $(2A_{1,1}+A).i$}& {\footnotesize $\{{\tilde
X}^3,{\tilde X}^3\}=-i{\tilde X}^1$}
&{\footnotesize $p\in\Re$}\\

\vspace{1mm}

&{\footnotesize $(2A_{1,1}+A).ii$}& {\footnotesize $\{{\tilde X}^3,{\tilde X}^3\}=i{\tilde X}^2$}&{\footnotesize
$p=\frac{1}{2}$}\\

\vspace{2mm}

&{\footnotesize $C^1_{-p}.i$ }&{\footnotesize $[{\tilde
X}^1,{\tilde X}^2]={\tilde X}^1,\;\;\;[{\tilde X}^2,{\tilde
X}^3]=p{\tilde X}^3$} &{\footnotesize $p\in\Re$}
\\

\vspace{2mm}

{\footnotesize $C^1_0$}&{\footnotesize
$C^1_{0,k}$}&{\footnotesize $[{\tilde X}^1,{\tilde X}^2]=k{\tilde
X}^2$}&{\footnotesize$k\in\Re-\{0\}$}
\\

\vspace{1mm}

{\footnotesize $C^1_{\frac{1}{2}}$}&{\footnotesize $I_{(2 , 1)} $}&\\

\vspace{1mm}

&{\footnotesize
$C^1_{p}.i{_{|_{p=-\frac{1}{2}}}}$}&{\footnotesize$[{\tilde
X}^1,{\tilde X}^2]= {\tilde X}^1,\;\;\;[{\tilde X}^2,{\tilde
X}^3]=\frac{1}{2} {\tilde X}^3$}&
 \\
\vspace{1mm}

&{\footnotesize
$C^1_{p}.ii{_{|_{p=-\frac{1}{2}}}}$}&{\footnotesize$[{\tilde
X}^1,{\tilde X}^2]=-{\tilde X}^1,\;\;\;[{\tilde X}^2,{\tilde
X}^3]=-\frac{1}{2} {\tilde X}^3$}&\\

\vspace{1mm}

&{\footnotesize $C^1_{\frac{1}{2}}.i$ }&{\footnotesize $[{\tilde
X}^1,{\tilde X}^2]={\tilde X}^1,\;\;\;[{\tilde X}^2,{\tilde
X}^3]=-\frac{1}{2} {\tilde X}^3,\;\;\; \{{\tilde X}^3,{\tilde
X}^3\}=i{\tilde X}^1$}&
\\

\vspace{1mm}

&{\footnotesize $C^1_{\frac{1}{2}}.ii$ }&{\footnotesize $[{\tilde
X}^1,{\tilde X}^2]=-{\tilde X}^1,\;\;\;[{\tilde X}^2,{\tilde
X}^3]=\frac{1}{2} {\tilde X}^3,\;\;\; \{{\tilde X}^3,{\tilde
X}^3\}=-i{\tilde X}^1$}&
\\

\vspace{1mm}

&{\footnotesize ${{C^1_{\frac{1}{2},k}}}$}&{\footnotesize
$[{\tilde X}^1,{\tilde X}^2]=k{\tilde X}^2,\;[{\tilde
X}^1,{\tilde X}^3]=\frac{k}{2}{\tilde X}^3,\;\{{\tilde
X}^3,{\tilde X}^3\}=ik{\tilde X}^2$}&{\footnotesize $k\in
{\Re-\{0\}}$}
\smallskip \\
\hline\hline

 \end{tabular}
\end{center}

%%%%%%%%%%%%%%%%%%%%%%%%%%%%%%%%%%%%%%%%%%%%%%%%%%%%%%%%%%%%%%%%%%%%%%%%%%%%%%%%%%%%%
\begin{center}
\hspace{4mm}{\small {\bf Table 5}} : \hspace{1mm}{ Two dimensional
Lie super-bialgebras of the type $(1 ,1)$.}\\
    \begin{tabular}{l l l l p{5mm} }
    \hline\hline
{\footnotesize ${\bf g}$ }& \hspace{1cm}{\footnotesize
$\tilde{\bf g}$
}&\hspace{2cm}{\footnotesize (Anti) Commutation relations of $\tilde{\bf g}$}\\
\hline

\vspace{2mm}

{\footnotesize $(A_{1,1}+A)$} &\hspace{0.9cm} {\footnotesize
$I_{(1,1)}$}&
\\

\vspace{1mm}

{\footnotesize $B$} & \hspace{1cm}{\footnotesize $I_{(1,1)}$}
\\

\vspace{1mm}

&\hspace{1cm}{\footnotesize $(A_{1,1}+A)$} &\hspace{2cm}
{\footnotesize $\{{\tilde X}^2,{\tilde X}^2\}=i{\tilde X}^1$}&
\\

\vspace{-1mm}

&\hspace{1cm}{\footnotesize $(A_{1,1}+A).i$} &
\hspace{2cm}{\footnotesize $\;\{{\tilde X}^2,{\tilde X}^2\}=-i
{\tilde X}^1$}&
\smallskip \\
\hline\hline
\end{tabular}
\end{center}

%%%%%%%%%%%%%%%%%%%%%%%%%%%%%%%%%%%%%%%%%%%%%%%%%%%%%%%%%%%%%%%%%%%%%%%%%%%%%%5555
%%%%%%%%%%%%%%%%%%%%%%%%%%%%%%%%%%%%%%%%%%%%%%%%%%%%%%%%%%%%%%%%%%%%%%%%%%%%%%%%%%%%%5555

\hspace{8mm}{\small {\bf Table 6}}: \hspace{1mm}{\small Three
dimensional  Lie super-bialgebras of the type (1 ,2). where
$\epsilon=
\pm 1$. }\\
   \begin{tabular}{l l l  p{0.5mm} }
    \hline\hline
{\footnotesize ${\bf g}$ }&  {\footnotesize $\tilde{\bf g}$} &
{\footnotesize Comments }\\ \hline
\vspace{1mm}
{\footnotesize $C^2_1$}&$I_{(1 , 2)} $&\\
\vspace{1mm}
 &{\footnotesize $(A_{1,1}+2A)^0_{0,0,\epsilon }$ } & &\\
\vspace{1mm}

&{\footnotesize $(A_{1,1}+2A)^1_{\epsilon,0,\epsilon}$ } & & \\
\vspace{2mm}

&{\footnotesize  $(A_{1,1}+2A)^2_{\epsilon,0,-\epsilon}$} & &
\\

\vspace{1mm}
{\footnotesize $C^2_p$}&{\footnotesize  $I_{(1 , 2)} $}&\\
\vspace{1mm}

{\footnotesize $-1 \leq p <1$ }& {\footnotesize
$(A_{1,1}+2A)^0_{\epsilon ,
0,0}\;,\;\;(A_{1,1}+2A)^0_{0,0,\epsilon}\;,\;\;(A_{1,1}+2A)^0_{\epsilon,\epsilon,\epsilon}$}
&& \\
\vspace{1mm}

&{\footnotesize $(A_{1,1}+2A)^1_{\epsilon , k,\epsilon }$}
 &{\footnotesize $ -1<k<1$}& \\

\vspace{2mm}

&{\footnotesize $(A_{1,1}+2A)^2_{0 , 1,0
}\;,\;\;(A_{1,1}+2A)^2_{\epsilon,1,0}\;,\;\;(A_{1,1}+2A)^2_{0,1,\epsilon}\;,\;\;
(A_{1,1}+2A)^2_{\epsilon ,k,-\epsilon }$ }&{\footnotesize $k
\in\Re$}&
\\

\vspace{1mm}
{\footnotesize $C^3$}&{\footnotesize $I_{(1 , 2)} $}&\\
\vspace{1mm}

&{\footnotesize $(A_{1,1}+2A)^0_{1 ,0,0}\;,\;\;(A_{1,1}+2A)^0_{0,0,1}$ }& \\
\vspace{1mm}

&{\footnotesize $(A_{1,1}+2A)^1_{\epsilon , 0,\epsilon }$}
 && \\

\vspace{2mm}

& {\footnotesize $(A_{1,1}+2A)^2_{0 ,\epsilon,0
}\;,\;\;(A_{1,1}+2A)^2_{\epsilon ,0,-\epsilon}$}
 && \\

\vspace{1mm}
{\footnotesize $C^4$}& {\footnotesize $I_{(1 , 2)} $}&\\
\vspace{1mm}

& {\footnotesize $(A_{1,1}+2A)^0_{\epsilon ,0,0}\;,\;\;(A_{1,1}+2A)^0_{0,0,\epsilon}$ }&&\\
\vspace{1mm}

& {\footnotesize $(A_{1,1}+2A)^1_{k ,
0,1}\;,\;\;(A_{1,1}+2A)^1_{s , 0,-1}$}
 &{\footnotesize $0<k,\;\;s<0$}& \\

\vspace{2mm}

&{\footnotesize $(A_{1,1}+2A)^2_{0 ,\epsilon,0
}\;,\;\;(A_{1,1}+2A)^2_{k ,0,1}\;,\;\;(A_{1,1}+2A)^2_{s,0,-1}$}
 &{\footnotesize $k<0,\;\;0<s$}\\

\vspace{1mm}
{\footnotesize $C^5_p$}&{\footnotesize $I_{(1 , 2)} $}&\\
\vspace{1mm}

{\footnotesize $p\geq 0$}&{\footnotesize $(A_{1,1}+2A)^0_{0,0,\epsilon}$ }&&\\
\vspace{1mm}

&{\footnotesize $(A_{1,1}+2A)^1_{k , 0,1}\;,\;\;(A_{1,1}+2A)^1_{s
, 0,-1}$} &{\footnotesize $0<k,\;\;s<0$}& \\

\vspace{2mm}

& {\footnotesize  $(A_{1,1}+2A)^2_{k
,0,1}\;,\;\;(A_{1,1}+2A)^2_{s,0,-1}$}
 &{\footnotesize $ k<0,\;\;0<s$}\smallskip\smallskip\\

\vspace{2mm}

{\footnotesize $(A_{1,1}+2A)^0$}&{\footnotesize  $I_{(1 , 2)} $}&\\

\vspace{2mm}

{\footnotesize $(A_{1,1}+2A)^1$}& {\footnotesize $I_{(1 , 2)} $}&\\

\vspace{-1mm}

{\footnotesize $(A_{1,1}+2A)^2$}&{\footnotesize  $I_{(1 , 2)} $}& \smallskip \\
\hline\hline
 \end{tabular}

%%%%%%%%%%%%%%%%%%%%%%%%%%%%%%%%%%%%%%%%%%%%%%%%%%%%%%%%%%%%%%%%%%%%%%%%%%%%%%%%%%
\vspace{3mm}

For three dimensional dual Lie superalgebras
$(A_{1,1}+2A)^0_{\alpha ,\beta,\gamma }\;,\;(A_{1,1}+2A)^1_{\alpha
,\beta,\gamma }$ and $(A_{1,1}+2A)^2_{\alpha ,\beta,\gamma }$
which are isomorphic with $(A_{1,1}+2A)^0,\;(A_{1,1}+2A)^1$ and
$(A_{1,1}+2A)^2$, respectively, we have the following
anticommutation relations:

\begin{equation}
\{{\tilde X}^2,{\tilde X}^2\}=i\alpha {\tilde X}^1 ,\quad
\{{\tilde X}^2,{\tilde X}^3\}=i \beta {\tilde X}^1 ,\quad
\{{\tilde X}^3,{\tilde X}^3\}=i\gamma {\tilde X}^1,\qquad \alpha,
\beta, \gamma \in\Re.
\end{equation}

Note that these Lie superalgebras are non isomorphic and they
differ in the bound of their parameters. The details of
differences between these lists with lists of Ref. \cite{H} are
given in appendix C. These differences cause the differences
between our theorem 3 with that of Ref. \cite{H}.

\vspace{7mm}

%%%%%%%%%%%%%%%%%%%%%%%%%%%%%%%%%%%%%%%%%%%%%%%%%%%%%%%%%%%%%%%%%%%%%%%%%%%%%%%%%%%%%
%%%%%%%%%%%%%%%%%%%%%%%%%%%%%%%%%%%%%%%%%%%%%%%%%%%%%%%%%%%%%%%%%%%%%%%%%%%%%%%%%%%%%

\section  {\large {\bf Four and six dimensional  Drinfel'd
superdoubles
}}
\smallskip
\goodbreak

Here we consider how many of four and six dimensional Lie
superalgebras $\cal{D}$ of  Drinfel'd superdoubles ${D}$ where
related respectively to 4 two dimensional and 70 three
dimensional \cite{R3} Lie super-bialgebras of tables 4-6 are
isomorphic. Indeed here we classify Lie superalgebras $\cal{D}$;
but for simply connected Lie supergroups $D$ this classification
is equivalent to the classification of Drinfel'd superdoubles $D$.
Furthermore for each Lie superalgebra $\cal{D}$ with Lie
super-bialgebras $(\bf g , \tilde {\bf g})$ there is other
decomposition $( \tilde {\bf g},\bf g )$ as their dual which must
be considered.  Now  we consider the Manin super triple $(\cal{D}
, {\bf g} , {\bf \tilde{g}})$ with the commutation relations (12)
and (13) for the Lie superalgebra $\cal{D}$. In general for
writing these commutation relations in the standard basis for
$\cal{D}$ we must first omit the $i$ coefficient from these
relations, then by use of $\{T_1,...,T_{m+\tilde m}\}$ bosonic
and  $\{T_{m+\tilde m+1},...,T_{n+\tilde n}\}$ fermionic basis
for ${\cal{D}}={\bf g}_{m+n}\oplus {\tilde {\bf g}}_{\tilde
m+\tilde n}$ we write the (anti)commutation relations for
$\cal{D}$ in the standard basis by multiply the $i$ to the
fermion-fermion anticommutation relations coefficients. In this
way we write the (anti)commutation relations of all four and six
dimensional Drinfel'd superdoubles [these are given in appendix
D]. Now we use the fact that two Lie superalgebras
$\cal{D}=\cal{D}_B+\cal{D}_F$ and
$\cal{D}^{'}={\cal{D}^{'}}_B+{\cal{D}^{'}}_F$ are isomorphic if
there are isomorphism between $\cal{D}_B \longrightarrow
{\cal{D}^{'}}_B$ and $\cal{D}_F \longrightarrow {\cal{D}^{'}}_F$.
This means that in the standard basis  there are block diagonal
isomorphism matrix $C$ between $\cal{D}$ and ${\cal{D}^{'}}$ such
that its superdeterminant [see appendix A for definition of
superdeterminant] is non zero  and satisfy in the following
relation \cite{ER1}
\begin{equation}
(-1)^{KL+MJ}\; C\; Y\hspace{.05mm}^M \;{C^{st}}
={Y'}\hspace{.05mm}^N \;C_N^{\;\;M},
\end{equation}
where $({Y}^M)_{\;KL} = -{{F}^{M}}_{KL}$ are the adjoint
representations and ${{F}^{M}}_{KL}$ are the structure constants
of Lie superalgebra $\cal{D}$ in the standard basis
$\{T_1,...,T_{n+\tilde n}\},\;(L, M, K=1,...,{n+\tilde n})$ and
the indices $K$ and $L$ correspond to the row and column of matrix
$Y\hspace{.05mm}^M$, respectively, and $J$ denotes the column of
matrix $C^{st}$ [here superscript ${st}$ stands for
supertranspose].  In this way we classify all four and six
dimensional Drinfeld's superdoubles. We perform this work by use
of MAPLE 10 program. The results are written as the following
three theorems. The isomorphism matrices are given in appendix B.
Note that the Drinfeld's superdoubles related for Manin super
triple $(\cal{D}$ , ${\bf g}_{(m,n)}$ , ${\bf \tilde{g}}_{({\tilde
m},{\tilde n})})$ are shown as $\cal{D}$$sd_{(m+\tilde m, n+\tilde
n)}$.

\vspace{3mm}

{\it Theorem 1:}$\;\;\;${\it  Every four dimensional Drinfeld's
superdoubles of the type $(2,2)$ belongs  to the one of the
following 3 classes and allows decomposition into all Lie
super-bialgebras listed in the class and their duals. }

\vspace{2mm}

\begin{tabular}{l l l l p{2mm} }

\vspace{1.5mm}
$\cal{D}$$sd^1_{(2,2)}:$&{\footnotesize $\Big(I_{(1 , 1)}, I_{(1 , 1)}\Big),$}&&\\

\vspace{1.5mm}

$\cal{D}$$sd^2_{(2,2)}:$&{\footnotesize $\Big((A_{1,1}+A), I_{(1 , 1)}\Big),$}&&\\

\vspace{1.5mm}

$\cal{D}$$sd^3_{(2,2)}:$&{\footnotesize $\Big(B, I_{(1 ,
1)}\Big),\;\;\Big(B,(A_{1,1}+A)\Big),\;\;
\Big(B,(A_{1,1}+A).i\Big)$.}&&\\

 \end{tabular}

\vspace{2mm}

{\it Theorem 2:}$\;\;\;${\it  Every six dimensional Drinfeld's
superdoubles of the type $(4,2)$ belongs to the one of the
following 9 classes and allows decomposition into all Lie
super-bialgebras listed in the class and their duals.  }

\vspace{2mm}

\begin{tabular}{l l l l p{2mm} }

\vspace{2mm}
$\cal{D}$$sd^1_{(4,2)}:$&{\footnotesize $\Big(I_{(2 , 1)}, I_{(2 , 1)}\Big),$}&&\\

\vspace{2mm}

$\cal{D}$$sd^2_{(4,2)}:$&{\footnotesize $\Big((2A_{1,1}+A), I_{(2 , 1)}\Big),$}&&\\

\vspace{2mm}

$\cal{D}$$sd^3_{(4,2)}:$&{\footnotesize $\Big((B+A_{1,1}), I_{(2
, 1)}\Big), \;\;\Big((B+A_{1,1}), (B+A_{1,1}).i\Big),\;
\Big(B,(2A_{1,1}+A)\Big),\;\Big(B,(2A_{1,1}+A).i\Big),$}&&\\

\vspace{2mm}

$\cal{D}$$sd^{4\;\;p=0}_{(4,2)}:$&{\footnotesize
$\Big(C^1_{p=0},I_{(2 ,
1)}\Big),\;\Big(C^1_{p=0},C^1_{-p=0}.i\Big),$}&&\\

\vspace{2mm}

$\cal{D}$$sd^{4\;\;p}_{(4,2)}:$&{\footnotesize $\Big(C^1_{p},I_{(2
, 1)}\Big),\;\Big (C^1_{-p},I_{(2 ,
1)}\Big),\;\Big(C^1_{p},C^1_{-p}.i\Big),\;
\Big(C^1_{p},(2A_{1,1}+A)\Big),\; \Big(C^1_{p},(2A_{1,1}+A).i\Big),
\;\;\;\;\;p\in\Re-\{0\},$}&&\\

\vspace{2mm}

$\cal{D}$$sd^{5}_{(4,2)}:$&{\footnotesize
$\Big(C^1_{p=0},(2A_{1,1}+A)\Big),\;
\Big(C^1_{p=0},(2A_{1,1}+A).i\Big),$}&&\\

\vspace{2mm}

$\cal{D}$$sd^{6\;\;k}_{(4,2)}:$&{\footnotesize $\Big(C^1_0,C^1_{0,k}\Big),\;\;k\in\Re-\{0\},$}&&\\

\vspace{2mm}

$\cal{D}$$sd^7_{(4,2)}:$&{\footnotesize
$\Big(C^1_{\frac{1}{2}},I_{(2 ,
1)}\Big),\;\Big(C^1_{p={\frac{1}{2}}},(2A_{1,1}+A).ii\Big),\;
\Big(C^1_{\frac{1}{2}},C^1_{p=-{\frac{1}{2}}}.i\Big),\;
\Big(C^1_{\frac{1}{2}},C^1_{p=-{\frac{1}{2}}}.ii\Big),
\;\Big(C^1_{\frac{1}{2}},C^1_{\frac{1}{2}}.i\Big),\;
\Big(C^1_{\frac{1}{2}},C^1_{\frac{1}{2}}.ii\Big),$}&&\\

\vspace{2mm}

$\cal{D}$$sd^{8\;\;k}_{(4,2)}:$&{\footnotesize
$\Big(C^1_{\frac{1}{2}},
{C^1_{\frac{1}{2}, k}} \Big),\;\;k\in\Re-\{0\}$.}&&\\
 \end{tabular}

\vspace{4mm}

\newpage

{\it Theorem 3:}$\;\;\;${\it Every six dimensional Drinfeld's
superdoubles of the type $(2,4)$ belongs  to the one of the
following 13 classes and allows decomposition into all  Lie
super-bialgebras listed in the class and their duals. }

\vspace{4mm}

\begin{tabular}{l l l l p{2mm} }

\vspace{4mm}

$\cal{D}$$sd^1_{(2,4)}:$&{\footnotesize $\Big(I_{(1 , 2)}, I_{(1 , 2)}\Big),$}&&\\

\vspace{2mm}

$\cal{D}$$sd^{2}_{(2,4)}:$&{\footnotesize $\Big(C^2_{0},
(A_{1,1}+2A)^0_{0,0,\epsilon_1}\Big),\;\Big(C^2_{0},
(A_{1,1}+2A)^0_{\epsilon_2,\epsilon_2,\epsilon_2}\Big),\;\Big(C^2_{0},
(A_{1,1}+2A)^1_{\epsilon_3,k,\epsilon_3}\Big),\;\Big(C^2_{0},
(A_{1,1}+2A)^2_{0,1,\epsilon_4}\Big),$}&&\\
\vspace{4mm} &{\footnotesize $\Big(C^2_{0},
(A_{1,1}+2A)^2_{\epsilon_5,s,-\epsilon_5}\Big),\;\;\;\;\;\;\;\;s\in
\Re,\;\;\;-1< k<1,\;\;\;\;\;\;
{\epsilon}_1,\cdots , {\epsilon}_5=\pm 1,$}&&\\

\vspace{5mm}

$\cal{D}$$sd^{3\;\;p=0}_{(2,4)}:$&{\footnotesize $\Big(C^2_{0},
(A_{1,1}+2A)^0_{\epsilon_1,0,0}\Big),\;\Big(C^2_{0},
(A_{1,1}+2A)^2_{0,1,0}\Big),\;\Big(C^2_{0},
(A_{1,1}+2A)^2_{\epsilon_2,1,0}\Big),\;\Big(C^2_{0}, I_{(1 ,
2)}\Big),\;\;\;\;\;\;\;\;
{\epsilon}_1, {\epsilon}_2=\pm 1,$}&&\\

\vspace{2mm}

$\cal{D}$$sd^{3\;\;p=1}_{(2,4)}:$ &{\footnotesize $\Big(C^2_{1},
(A_{1,1}+2A)^0_{0,0,\epsilon_1}\Big),\;\Big(C^2_{1},
(A_{1,1}+2A)^1_{\epsilon_2,0,\epsilon_2}\Big),\;\Big(C^2_{1},
(A_{1,1}+2A)^2_{\epsilon_3,0,-\epsilon_3}\Big),\;\Big(C^2_{p=-1},
(A_{1,1}+2A)^0_{\epsilon_4,0,0}\Big),\;$}&&\\
\vspace{2mm} &{\footnotesize $\Big(C^2_{p=-1},
(A_{1,1}+2A)^0_{0,0,\epsilon_5}\Big),\;\Big(C^2_{p=-1},
(A_{1,1}+2A)^1_{\epsilon_6,k=0,\epsilon_6}\Big),\;\Big(C^2_{p=-1},
(A_{1,1}+2A)^2_{\epsilon_7,s=0,-\epsilon_7}\Big),\;\Big(C^2_{1},
I_{(1 , 2)}\Big),$}&&\\

\vspace{5mm} &{\footnotesize $\Big(C^2_{p=-1}, I_{(1 ,
2)}\Big),\;\;\;\;\;\;\;\;
{\epsilon}_1,\cdots , {\epsilon}_7=\pm 1,$}&&\\

\vspace{2mm}

$\cal{D}$$sd^{3\;\;p}_{(2,4)}:$ &{\footnotesize $\Big(C^2_{p},
(A_{1,1}+2A)^0_{\epsilon_1,0,0}\Big),\;\Big(C^2_{-p},
(A_{1,1}+2A)^0_{\epsilon_2,0,0}\Big),\;\Big(C^2_{p},
(A_{1,1}+2A)^0_{0,0,\epsilon_3}\Big),\;\Big(C^2_{p},
(A_{1,1}+2A)^0_{\epsilon_4,\epsilon_4,\epsilon_4}\Big),$}&&\\
\vspace{2mm}

&{\footnotesize $\Big(C^2_{p},
(A_{1,1}+2A)^1_{\epsilon_5,k,\epsilon_5}\Big),\;\Big(C^2_{p},
(A_{1,1}+2A)^2_{0,1,0}\Big),\;\Big(C^2_{p},
(A_{1,1}+2A)^2_{\epsilon_{6},1,0}\Big),\;\Big(C^2_{p},
(A_{1,1}+2A)^2_{0,1,\epsilon_{7}}\Big)$}&&\\

\vspace{2mm}

&{\footnotesize $\Big(C^2_{p},
(A_{1,1}+2A)^2_{\epsilon_{8},s,-\epsilon_{8}}\Big),\;\Big(C^2_{p},
I_{(1 , 2)}\Big),\;\;\;\;\;\;p \in (-1,1)-\{0\},$}&&\\

\vspace{5mm} &{\footnotesize $\;\;\;\;\;s\in \Re,\;\;\;-1<
k<1,\;\;\;\;\;{\epsilon}_1,\cdots ,
{\epsilon}_{8}=\pm 1$}, & &\\

\vspace{5mm}

$\cal{D}$$sd^4_{(2,4)}:$&{\footnotesize $\Big(C^3,
(A_{1,1}+2A)^0_{1,0,0}\Big),\;\Big(C^3,
(A_{1,1}+2A)^2_{0,\epsilon,0}\Big),\;\Big(C^3, I_{(1 , 2)}\Big),\;
\Big( (A_{1,1}+2A)^2, I_{(1 , 2)}\Big),\;\;\;\;\;\;\;\;
{\epsilon}=\pm 1,$}&&\\

\vspace{5mm}

$\cal{D}$$sd^5_{(2,4)}:$&{\footnotesize $\Big(C^3,
(A_{1,1}+2A)^0_{0,0,1}\Big),\; \Big(C^3,
(A_{1,1}+2A)^1_{{\epsilon}_1,0,{\epsilon}_1}\Big),\;\Big(C^3,
(A_{1,1}+2A)^2_{{\epsilon}_2,0,-{\epsilon}_2}\Big),\;\;\;
\;\;\;\;\;\;\;\;
{\epsilon}_1, {\epsilon}_2=\pm 1,$}&&\\

\vspace{2mm}

$\cal{D}$$sd^6_{(2,4)}:$&{\footnotesize $\Big(C^4,
(A_{1,1}+2A)^0_{{\epsilon}_1,0,0}\Big),\; \Big(C^4,
(A_{1,1}+2A)^0_{0,0,{\epsilon}_2}\Big),\;\Big(C^4,
(A_{1,1}+2A)^1_{m,0,{\epsilon}_3}\Big),\;\Big(C^4,
(A_{1,1}+2A)^2_{n,0,{\epsilon}_4}\Big),$}&&\\

\vspace{2mm} &{\footnotesize $\Big(C^4,
(A_{1,1}+2A)^2_{0,{\epsilon}_5,0}\Big),\;\Big(C^2_{p=-1},
(A_{1,1}+2A)^0_{\epsilon_6,\epsilon_6,\epsilon_6}\Big),\;\Big(C^2_{p=-1},
(A_{1,1}+2A)^1_{\epsilon_7,k,\epsilon_7}\Big),\;\Big(C^2_{p=-1},
(A_{1,1}+2A)^2_{0,1,0}\Big),$}&&\\

\vspace{2mm} &{\footnotesize $\Big(C^2_{p=-1},
(A_{1,1}+2A)^2_{\epsilon_8,1,0}\Big),\;\Big(C^2_{p=-1},
(A_{1,1}+2A)^2_{0,1,\epsilon_9}\Big),\;\Big(C^2_{p=-1},
(A_{1,1}+2A)^2_{\epsilon_{10},s,-\epsilon_{10}}\Big),\;\Big(C^4,
I_{(1 , 2)}\Big),$}&&\\

\vspace{0.5mm}

&{\footnotesize $\;\;\;\;\;k \in (-1 , 1)-\{0\},\;\;\;s \in
\Re-\{0\},\;\;\;\;\;
{\epsilon}_1,\cdots , {\epsilon}_{10}=\pm 1,$}&&\\

\vspace{0.5mm}

&{\footnotesize
$\;\;\;\;m>0\;\;\;\;if\;\epsilon_3=1;\;\;\;\;\;\;\;\;\;\;\;\;\;\;n>0\;
\;\;\;if\;\epsilon_4=-1;$}&&\\

\vspace{4mm}

&{\footnotesize
$\;\;\;\;m<0\;\;\;\;if\;\epsilon_3=-1,\;\;\;\;\;\;\;\;\;\;\;n<0\;
\;\;\;if\;\epsilon_4=1,$}&&\\

\vspace{1mm}

$\cal{D}$$sd^{7\;\;p=0}_{(2,4)}:$&{\footnotesize $\Big(C^5_{0},
(A_{1,1}+2A)^0_{0,0,\epsilon_1}\Big),\;\Big(C^5_{0},
(A_{1,1}+2A)^1_{k,0,\epsilon_2}\Big),\;\Big(C^5_{0},
(A_{1,1}+2A)^2_{s,0,\epsilon_3}\Big),$}&&\\

\vspace{0.5mm}

&{\footnotesize
$\;\;\;\;k>0\;\;\;\;if\;\epsilon_2=1;\;\;\;\;\;\;\;\;\;\;\;\;\;\;s>0,\;\;
s\neq 1\; \;\;\;if\;\epsilon_3=-1;$}&&\\

\vspace{5mm}

&{\footnotesize
$\;\;\;\;k<0\;\;\;\;if\;\epsilon_2=-1,\;\;\;\;\;\;\;\;\;\;\;s<0,\;\;
s\neq -1\;
\;\;\;if\;\epsilon_3=1,$}&&\\

\vspace{1mm}

$\cal{D}$$sd^{7\;\;p}_{(2,4)}:$&{\footnotesize $\Big(C^5_{p},
(A_{1,1}+2A)^0_{0,0,\epsilon_1}\Big),\;\Big(C^5_{p},
(A_{1,1}+2A)^1_{k,0,\epsilon_2}\Big),\;\Big(C^5_{p},
(A_{1,1}+2A)^2_{s,0,\epsilon_3}\Big),\;\Big(C^5_{p},
I_{(1,2)}\Big),\;\;p>0,$}&&\\

\vspace{0.5mm}

&{\footnotesize
$\;\;\;\;k>0\;\;\;\;if\;\epsilon_2=1;\;\;\;\;\;\;\;\;\;\;\;\;\;\;s>0\;
\;\;\;if\;\epsilon_3=-1;$}&&\\

\vspace{4mm}

&{\footnotesize
$\;\;\;\;k<0\;\;\;\;if\;\epsilon_2=-1,\;\;\;\;\;\;\;\;\;\;\;s<0\;
\;\;\;if\;\epsilon_3=1,$}&&\\

\vspace{4mm}

$\cal{D}$$sd^{8}_{(2,4)}:$&{\footnotesize $\Big(C^5_{0},
(A_{1,1}+2A)^2_{1,0,-1}\Big),\;\Big(C^5_{0},
(A_{1,1}+2A)^2_{-1,0,1}\Big),\;\Big(C^5_{0},
I_{(1,2)}\Big),$}&&\\

\vspace{4mm}

$\cal{D}$$sd^{9}_{(2,4)}:$&{\footnotesize $\Big((A_{1,1}+2A)^0 ,
I_{(1,2)} \Big),$}&&\\

$\cal{D}$$sd^{10}_{(2,4)}:$&{\footnotesize $\Big((A_{1,1}+2A)^1 ,
I_{(1,2)} \Big).$}&&\\
\end{tabular}

\newpage
%%%%%%%%%%%%%%%%%%%%%%%%%%%%%%%%%%%%%%%%%%%%%%%%%%%%%%%%%%%%%%%%%%%%%%%%%%%%%%%%%%%%%
\section  {\large {\bf Conclusion}}

We classify  two and three dimensional Lie super-bialgebras
obtained from decomposable Lie superalgebras. We also classified
all four and six dimensional Drinfel'd superdoubles as three
theorems. Using this classification one can investigate
Poisson-Lie T plurality of sigma models over Lie supergroups.
Obtaining the modular spaces of these Drinfel'd superdoubles is
the other open problem.

\bigskip
{\bf Acknowledgments}

\vspace{3mm} We would like to thank  F. Darabi for carefully
reading the manuscript and useful comments.

\vspace{5mm}
%%%%%%%%%%%%%%%%%%%%%%%%%%%%%%%%%%%%%%%%%%%%%%%%%%%%%%%%%%%%%%%%%%%%%%%%%%%%%%%%%%%%%
{\bf Appendix A:  Some notations about supermatrices and
supertensors }

\vspace{3mm}

In this appendix,  We consider the standard basis for the
supervector spaces so that in writing the basis as a column
matrix, we first present the bosonic base and then the fermionic
one. The transformation of standard basis and its dual basis can
be written as follows:
$$
 {e'}_i=(-1)^j{K_i}\;^j e_j ,\hspace{10mm}{e'}^i={{K^{-{st}}}^i}_j\; e^j,
$$

where the transformation matrix $K$ has the following block
diagonal representation \cite{D}
\begin{equation}
K=\left( \begin{tabular}{c|c}
                 A & C \\ \hline
                 D & B \\
                 \end{tabular} \right),
\end{equation} where $A,B$ and $C$ are real submatrices and $D$ is pure
imaginary submatrix \cite{D}. Here we consider the matrix and
tensors having a form with all upper and lower indices written in
the right hand side.

\bigskip

1. The transformation properties of upper and lower right indices
to the left one for general tensors are as follows:
\begin{equation}
^iT_{jl...}^{\;k}=T_{jl...}^{ik},\qquad
_jT^{ik}_{l...}=(-1)^j\;T_{jl...}^{ik}.
\end{equation}

\bigskip

2. For supertransposition we have
$$
{{L}^{{st}\;i}}_j=(-1)^{ij}\;{L_j}^{\;i},\qquad
{L^{st}_{\;i}}^{\;j}=(-1)^{ij}\;{L^i}_{\;j},
$$
\begin{equation}
M^{st}_{\;ij}=(-1)^{ij}\;M_{\;ji},\qquad
M^{{st}\;ij}=(-1)^{ij}\;M^{\;ji}.
\end{equation}

\bigskip

3. For superdeterminant we have
\begin{equation}
sdet\left( \begin{tabular}{c|c}
                 A & C \\ \hline
                 D & B \\
                 \end{tabular} \right)=det{(A-CB^{-1}D)}(det B)^{-1},
\end{equation}
when det $B\neq0$ and
\begin{equation}
sdet\left( \begin{tabular}{c|c}
                 A & C \\ \hline
                 D & B \\
                 \end{tabular} \right)=(det{(B-DA^{-1}C)})^{-1}\;(det A),
\end{equation}
when det $A\neq0$.

\vspace{8mm}

%%%%%%%%%%%%%%%%%%%%%%%%%%%%%%%%%%%%%%%%%%%%%%%%%%%%%%%%%%%%%%%%%%%%%%%%%%%%%%%%%%%%%%%%%%%%%%%%%%%%%%%%%%%%%%%%%%%%%
%%%%%%%%%%%%%%%%%%%%%%%%%%%%%%%%%%%%%%%%%%%%%%%%%%%%%%%%%%%%%%%%%%%%%%%%%%%%%%%%%%%%%
%%%%%%%%%%%%%%%%%%%%%%%%%%%%%%%%%%%%%%%%%%%%%%%%%%%%%%%%%%%%%%%%%%%%%%%%%%%%%%%%%%%%%%%%%%%%%%%%%%%%%%%%%%%%%%%%%%%%%
%%%%%%%%%%%%%%%%%%%%%%%%%%%%%%%%%%%%%%%%%%%%%%%%%%%%%%%%%%%%%%%%%%%%%%%%%%%%%%%%%%%%%
{\bf Appendix B:  Solutions of super Jacobi and mixed super Jacobi
identities $(sJ-msJ)$ for dual of decomposable Lie superalgebras
and isomorphism matrices $C$}

\bigskip

1. Solutions of $(sJ-msJ)$ for dual  Lie superalgebras of
$(B+A_{1,1})$ are

\vspace{2mm}

\begin{tabular}{l l l l p{2mm} }

\vspace{2mm}

$i)$&${{\tilde{f}}^{23}}_{\; \; \:
3}=\alpha,$&$\;\;\;\;\;\;\;\;\;\alpha\in\Re,$&\\

$ii)$ &${{\tilde{f}}^{33}}_{\; \; \:
1}=i\beta,$ &$\;\;\;\;\;\;\;\;\;\beta \in\Re.$&\\

\end {tabular}

\smallskip

Isomorphism matrix $C$ between  solution $(i)$  and  $(B+A_{1,1})$
is as follows:

$$
C=\left( \begin{tabular}{ccc}
              $ c_{11}$ & $c_{12}$& $ c_{13}$ \\
                $ c_{21}$ & 0 & 0 \\
                 0  & 0 & $ c_{33}$ \\
                 \end{tabular}
                 \right),\;\;\;\;c_{12},c_{21},c_{33}\in\Re-\{0\};\;\;c_{11}\in\Re,
$$

where $c_{12}=\frac{1}{{{\tilde{f}}^{23}}_{\; \; \: 3}}$ and by
imposing that $C$ must be the transformation matrix (18), we then
have $c_{13}=0$.

\smallskip

For solution $(ii)$ we have isomorphism matrix between this
solution and  $(2A_{1,1}+A)$ as follows:

$$
C=\left( \begin{tabular}{ccc}
              $ c_{11}$ & $0$& $ 0$ \\
                $ c_{21}$ & $c_{22}$ & 0 \\
                 $c_{31}$ &  $c_{32}$ & $ c_{33}$ \\
                 \end{tabular}
                 \right),\;\;\;\;c_{11},c_{22},c_{33}\in\Re-\{0\};\;\;c_{21}\in\Re,
$$

where $c_{11}=-i{c_{33}}^2 {{\tilde{f}}^{33}}_{\; \; \: 1}$ and by
imposing that $C$ must be the transformation matrix, we then have
$c_{31}, c_{32}=0$.

\bigskip
%%%%%%%%%%%%%%%%%%%%%%%%%%%%%%%%%%%%%%%%%%%%%%%%%%%%%%%%%%%%%%%%%%%%%%%%%%%%
2. Solutions of $(sJ-msJ)$ for dual  Lie superalgebras  of $C^1_0$
are

\vspace{3mm}

\begin{tabular}{l l l l p{2mm} }

\vspace{2mm}

$i)$&${{\tilde{f}}^{12}}_{\; \; \:
1}=\alpha,\;\;{{\tilde{f}}^{12}}_{\; \; \:
2}=\beta,$&$\;\;\;\;\;\;\;\alpha, \beta \in\Re,$&\\

\vspace{2mm}

$ii)$&${{\tilde{f}}^{12}}_{\; \; \:
1}=\gamma,$&$\;\;\;\;\;\;\;\gamma \in\Re,$&\\

$iii)$ &${{\tilde{f}}^{33}}_{\; \; \:
1}=i\delta,$ &$\;\;\;\;\;\;\;\;\delta \in\Re.$&\\

\end {tabular}

\smallskip

Isomorphism matrix $C$ between solution $(i)$ and  $(C^1_0)$ is as
follows:

$$
C=\left( \begin{tabular}{ccc}
              $ c_{11}$ & $c_{12}$& $ c_{13}$ \\
                $ c_{21}$ & $c_{22}$ & 0 \\
                 $0$ &  $0$ & $ c_{33}$ \\
                 \end{tabular}
                 \right),\;\;\;\;\;c_{33}\in\Re-\{0\};\;\;c_{11}c_{22} \neq c_{21} c_{12},
$$

where $c_{21}=\frac{{{\tilde{f}}^{12}}_{\; \; \:
1}}{{{\tilde{f}}^{12}}_{\; \; \: 2}}{c_{22}},
c_{12}=\frac{c_{11}{{\tilde{f}}^{12}}_{\; \; \:
2}-1}{{{\tilde{f}}^{12}}_{\; \; \: 1}}$ and  by imposing that $C$
must be the transformation matrix, we then have $c_{13}=0$.
Isomorphism matrix in the solution (ii) is a special case of the
above isomorphism matrix and for solution (iii) we have $sdetC=0$.

\bigskip
%%%%%%%%%%%%%%%%%%%%%%%%%%%%%%%%%%%%%%%%%%%%%%%%%%%%%%%%%%%%%%%%%%%%%%%%%%%%%%%%%%%%%%%%%
3. Solutions of $(sJ-msJ)$ for  dual Lie superalgebras of
$C^1_p\;(p\in\Re)$ are

\vspace{3mm}

\begin{tabular}{l l l l p{2mm} }

\vspace{2mm}

$i)$&${{\tilde{f}}^{12}}_{\; \; \:
1}=\alpha,\;\;{{\tilde{f}}^{23}}_{\; \; \:
3}=p\alpha,$&$\;\;\;\;\;\;\;\;p\in\Re,\;\;\;\alpha \in\Re,$&\\

\vspace{2mm}

$ii)$&${{\tilde{f}}^{12}}_{\; \; \: 1}={{\tilde{f}}^{23}}_{\; \;
\: 3}=\beta,$&$\;\;\;\;\;\;\;\;p=1,\;\;\;\beta \in\Re,$&\\

\vspace{2mm}

$iii)$ &${{\tilde{f}}^{33}}_{\; \; \:
1}=i\gamma,$ &$\;\;\;\;\;\;\;\;p\in\Re,\;\;\;\gamma \in\Re,$&\\

\vspace{2mm}

$iv)$ &${{\tilde{f}}^{33}}_{\; \; \:
1}=i\lambda,\;\;{{\tilde{f}}^{33}}_{\; \; \:
2}=i\eta$ &$\;\;\;\;\;\;\;\;p=\frac{1}{2},\;\;\;\lambda, \eta \in\Re,$&\\

\vspace{2mm}

$v)$ &${{\tilde{f}}^{12}}_{\; \; \:
1}=\mu,\;\;{{\tilde{f}}^{23}}_{\; \; \:
3}=-\frac{\mu}{2},\;\;{{\tilde{f}}^{33}}_{\; \; \:
1}=i\nu$ &$\;\;\;\;\;\;\;\;p=-\frac{1}{2},\;\;\;\mu, \nu \in\Re.$&\\

\end {tabular}

\smallskip

For solutions $i)$ and $ii)$ we have $sdet C=0$. For solution
$iii)$ isomorphism matrix between this solution and
$(2A_{1,1}+A)$ is as follows:

$$
C=\left( \begin{tabular}{ccc}
              $ c_{11}$ & $0$& $ 0$ \\
                $ c_{21}$ & $c_{22}$ & 0 \\
                 $c_{31}$ &  $c_{32}$ & $ c_{33}$ \\
                 \end{tabular}
                 \right),\;\;\;\;c_{11},c_{22},c_{33}\in\Re-\{0\};\;\;c_{21}\in\Re,
$$
where $c_{11}=-i{c_{33}}^2 {{\tilde{f}}^{33}}_{\; \; \: 1}$ and by
imposing that $C$ must be the transformation matrix, we then have
$c_{31}, c_{32}=0$.

\smallskip

For solution $iv)$  isomorphism matrix  between this solution and
$(2A_{1,1}+A)$ is as follows:

$$
C=\left( \begin{tabular}{ccc}
              $ c_{11}$ & $c_{12}$& $ 0$ \\
                $ c_{21}$ & $c_{22}$ & 0 \\
                 $c_{31}$ &  $c_{32}$ & $ c_{33}$ \\
                 \end{tabular}
                 \right),\;\;\;\;\;c_{33}\in\Re-\{0\};\;\;c_{11}c_{22}\neq c_{21} c_{21},
$$
where $c_{11}=-i{c_{33}}^2 {{\tilde{f}}^{33}}_{\; \; \: 1},\;
c_{12}=-i{c_{33}}^2 {{\tilde{f}}^{33}}_{\; \; \: 2}$ and by
imposing that $C$ must be the transformation matrix, we then have
$c_{31}, c_{32}=0$.

\smallskip

For solution $v)$  isomorphism matrix  between $(C^1_p ,
(2A_{1,1}+A))$ is  the same of isomorphism matrix in the solution
iii) with the same of conditions.

\bigskip

%%%%%%%%%%%%%%%%%%%%%%%%%%%%%%%%%%%%%%%%%%%%%%%%%%%%%%%%%%%%%%%%%%%%%%%%%%%%%%%%%%%%%%%%%
4. Solutions of $(sJ-msJ)$ for  dual Lie superalgebras of
$C^2_1,\; C^2_p\;(-1\leq p< 1), \;C^3, C^4$ and $C^5_p\;(p \geq
0)$ are

\vspace{2mm}

\begin{tabular}{l l l l p{2mm} }

\vspace{2mm}

&${{\tilde{f}}^{22}}_{\; \; \:
1}=i\alpha,\;\;{{\tilde{f}}^{23}}_{\; \; \: 1}=i
\beta,\;\;{{\tilde{f}}^{33}}_{\; \; \:
1}=i \gamma,$&$\;\;\;\;\;\;\;\;\alpha, \beta, \gamma \in\Re.$&\\

\end {tabular}

\smallskip

For the above set of solutions  we have the following isomorphism
matrices which map  $C^2_1, C^2_p, C^3, C^4$ and $C^5_p$ into
$(A_{1,1}+2A)^0$

$$
C_1=\left( \begin{tabular}{ccc}
              $ c_{11}$ & $0$& $ 0$ \\
                $ c_{21}$ & $c_{22}$ & $c_{23}$ \\
                 $c_{31}$ &  $0$ & $ c_{33}$ \\
                 \end{tabular}
                 \right),\;\;\;\;c_{11},c_{22},c_{33}\in\Re-\{0\};\;\;c_{23}\in\Re,
$$
where $c_{11}=-i{c_{22}}^2 {{\tilde{f}}^{22}}_{\; \; \: 1}$ and
${{\tilde{f}}^{22}}_{\; \; \: 1}={{\tilde{f}}^{33}}_{\; \; \:
1}=0$.

\smallskip

$$
C_2=\left( \begin{tabular}{ccc}
              $ c_{11}$ & $0$& $ 0$ \\
                $ c_{21}$ & $c_{22}$ & $c_{23}$ \\
                 $c_{31}$ &  $c_{32}$ & $ c_{33}$ \\
                 \end{tabular}
                 \right),\;\;\;\;\;c_{11}\in\Re-\{0\};\;\;c_{22}c_{33}\neq c_{23} c_{32},
$$
where $ {{\tilde{f}}^{33}}_{\; \; \: 1}=
\frac{ic_{11}{c_{32}}^2}{(c_{22}c_{33}-c_{23} c_{32})^2},\;
{{\tilde{f}}^{23}}_{\;\;\:1}=-\frac{c_{33}}{c_{32}}{{\tilde{f}}^{33}}_{\;
\; \: 1}$ and
${{\tilde{f}}^{22}}_{\;\;\:1}=(\frac{c_{33}}{c_{32}})^2
{{\tilde{f}}^{33}}_{\; \; \: 1}$.

\smallskip

$$
C_3=\left( \begin{tabular}{ccc}
              $ c_{11}$ & $0$& $ 0$ \\
                $ c_{21}$ & $c_{22}$ & $c_{23}$ \\
                 $c_{31}$ &  $c_{32}$ & $ 0$ \\
                 \end{tabular}
                 \right),\;\;\;\;\;c_{11}, c_{23}, c_{32}\in\Re-\{0\};\;\;c_{22}\in \Re,
$$
where $ {{\tilde{f}}^{33}}_{\; \; \: 1}=
\frac{ic_{11}}{{c_{23}}^2},\;
{{\tilde{f}}^{23}}_{\;\;\:1}={{\tilde{f}}^{22}}_{\;\;\:1}=0$.

\smallskip

Imposing that $C_1, C_2$ and $C_3$ must be the transformation
matrices, we have $c_{21}, c_{31}=0$.

\vspace{6mm}
%%%%%%%%%%%%%%%%%%%%%%%%%%%%%%%%%%%%%%%%%%%%%%%%%%%%%%%%%%%%%%%%%%%%%%%%%%%%%%%%%%%%%
{\bf Appendix C: The details of differences  between the lists of
Lie super-bialgebras of Ref. \cite{ER1} with Ref. \cite{H}}

\begin{center}
\begin{tabular}{l l l l  l l p{15mm} }
    \hline\hline

\vspace{1mm}

{\footnotesize ${\bf g}$ }& {\footnotesize $\tilde {\bf g}$ of
Ref. \cite{ER1} }&  {\footnotesize $\tilde {\bf g}$ of Ref.
\cite{H} } & \\\hline

\vspace{1mm}

{\footnotesize $C^2_p$ }& {\footnotesize   $I_{(1,2)}$} &
{\footnotesize $A_{12}$}&
\\

\vspace{1mm}

{\footnotesize $-1\leq p <1$ }&{\footnotesize
$(A_{1,1}+2A)^0_{\epsilon,0,0}$} & {\footnotesize
${N^{\epsilon,\delta,0}_{12}}_{|_{\delta=0}}$}&
\\

\vspace{1mm}

&{\footnotesize  $(A_{1,1}+2A)^0_{0,0,\epsilon}$} &
{\footnotesize ${{N^{0,\delta,\epsilon}_{12}}_{|_{\delta=0}}}$}&
\\

\vspace{1mm}

&{\footnotesize  $(A_{1,1}+2A)^0_{1,1,1}$} & {\footnotesize
${{N^{\epsilon_1,k,\epsilon_2}_{12}}_{|_{\epsilon_1=\epsilon_2=1\atop{k=1}}}}$}&
\\

\vspace{1mm}

&{\footnotesize $(A_{1,1}+2A)^0_{-1,-1,-1}$} & {\footnotesize Not
exist}&
\\

\vspace{1mm}

&{\footnotesize
${(A_{1,1}+2A)^1_{\epsilon,k,\epsilon}}_{|_{0<k<1}}$} &
{\footnotesize
${{N^{\epsilon_1,k,\epsilon_2}_{12}}_{|_{\epsilon_1=\epsilon_2\atop{0<k<1}}}}$}&
\\

\vspace{1mm}

&{\footnotesize
${(A_{1,1}+2A)^1_{\epsilon,k,\epsilon}}_{|_{-1<k<0}}$} &
{\footnotesize Not exist}&
\\

\vspace{1mm}

&{\footnotesize Not exist} & {\footnotesize
${{N^{\epsilon_1,k,\epsilon_2}_{12}}_{|_{\epsilon_1=\epsilon_2\atop{k
\geq 1}}}}$}&
\\

\vspace{1mm}

&{\footnotesize ${(A_{1,1}+2A)^2_{\epsilon,k,-\epsilon}}_{|_{k
\geq 0 }}$} & {\footnotesize
${{N^{\epsilon_1,k,\epsilon_2}_{12}}_{|_{\epsilon_1=-\epsilon_2\atop{k
\geq 0 }}}}$}&
\\

\vspace{1mm}

&{\footnotesize
${(A_{1,1}+2A)^2_{\epsilon,k,-\epsilon}}_{|_{k<0}}$} &
{\footnotesize Not exist}&
\\

\vspace{1mm}

&{\footnotesize ${(A_{1,1}+2A)^2_{0,1,0}}$} & {\footnotesize
${{N^{0,1,0}_{12}}}$}&
\\

\vspace{1mm}

&{\footnotesize ${(A_{1,1}+2A)^2_{\epsilon,1,0}}$} &
{\footnotesize ${N^{\epsilon,\delta,0}_{12}}_{|_{\delta=1}}$}&
\\

\vspace{1mm}

&{\footnotesize ${(A_{1,1}+2A)^2_{0,1,\epsilon}}$} &
{\footnotesize ${N^{0,\delta,\epsilon}_{12}}_{|_{\delta=1}}$}&
\smallskip \\\hline

\vspace{1mm}

    \end{tabular}
\end{center}
%%%%%%%%%%%%%%%%%%%%%%%%%%%%%%%%%%%%%%%%%%%%%%%%%%%%%%%%%%%%%%%%%%%%%%%%%%%%%%%%%%%%%%%555
\begin{center}
\hspace{1cm}{\small Continued }\smallskip\\
\begin{tabular}{l l l l  l l p{15mm} }
 \hline

\vspace{1mm}

{\footnotesize ${\bf g}$ }& {\footnotesize $\tilde {\bf g}$ of
Ref. \cite{ER1} }&  {\footnotesize $\tilde {\bf g}$ of Ref.
\cite{H} } & \\\hline

\vspace{1mm}

{\footnotesize   $C^2_1$}& {\footnotesize   $I_{(1,2)}$} &
{\footnotesize $A_{12}$}&
\\

\vspace{1mm}

& {\footnotesize   $(A_{1,1}+2A)^0_{0,0,\epsilon}$} &
{\footnotesize $N^{0,0,\epsilon}_{12}$}&
\\

\vspace{1mm}

& {\footnotesize $(A_{1,1}+2A)^1_{\epsilon,0,\epsilon}$} &
{\footnotesize $N^{\epsilon,0,\epsilon}_{12}$}&
\\

\vspace{1mm}

&{\footnotesize   $(A_{1,1}+2A)^2_{1,0,-1}$} & {\footnotesize
$N^{1,0,-1}_{12}$}&
\\

\vspace{2mm}

& {\footnotesize   $(A_{1,1}+2A)^2_{-1,0,1}$} & {\footnotesize
Not exist}&
\smallskip \\

%%%%%%%%%%%%%%%%%%%%%%%%%%%%%%%%%%%%%%%%%%%%%%%%%%%%%%%%%%%%%%%%%%%%
{\footnotesize $C^3$ }& {\footnotesize   $I_{(1,2)}$} &
{\footnotesize $A_{12}$}&
\\

\vspace{1mm}

&{\footnotesize  $(A_{1,1}+2A)^0_{1,0,0}$} & {\footnotesize
${N^{1,0,0}_{12}}$}&
\\

\vspace{1mm}

&{\footnotesize  $(A_{1,1}+2A)^0_{0,0,1}$} & {\footnotesize
${{N^{0,0,1}_{12}}}$}&
\\

\vspace{1mm}

&{\footnotesize  $(A_{1,1}+2A)^1_{1,0,1}$} & {\footnotesize
${{N^{\epsilon,0,1}_{12}}_{|_{\epsilon=1}}}$}&
\\

\vspace{1mm}

&{\footnotesize ${(A_{1,1}+2A)^1_{-1,0,-1}}$} & {\footnotesize
Not exist}&
\\

\vspace{1mm}

&{\footnotesize $(A_{1,1}+2A)^2_{0,\epsilon,0}$} & {\footnotesize
${N^{0,\epsilon,0}_{12}}$}&
\\

\vspace{1mm}

&{\footnotesize ${(A_{1,1}+2A)^2_{-1,0,1}}$} & {\footnotesize
${{N^{\epsilon,0,1}_{12}}_{|_{\epsilon=-1}}}$}&
\\

\vspace{2mm}

&{\footnotesize ${(A_{1,1}+2A)^2_{1,0,-1}}$} & {\footnotesize Not
exist}&\smallskip \\

%%%%%%%%%%%%%%%%%%%%%%%%%%%%%%%%%%%%%%%%%%%%%%%%%%%%%%%%%%%%%%%%%%%%%%%%%%%%%%%
{\footnotesize $C^4$ }& {\footnotesize   $I_{(1,2)}$} &
{\footnotesize $A_{12}$}&
\\

\vspace{1mm}

&{\footnotesize  $(A_{1,1}+2A)^0_{\epsilon,0,0}$} & {\footnotesize
${N^{\epsilon,0,0}_{12}}$}&
\\

\vspace{1mm}

&{\footnotesize  $(A_{1,1}+2A)^0_{0,0,\epsilon}$} & {\footnotesize
${{N^{k,0,\epsilon}_{12}}_{|_{k=o}}}$}&
\\

\vspace{1mm}

&{\footnotesize  $(A_{1,1}+2A)^1_{k,0,1}$} & {\footnotesize
${{N^{k,0,\epsilon}_{12}}_{|_{\epsilon=1\atop{k>0}}}}$}&
\\

\vspace{1mm}

&{\footnotesize  $(A_{1,1}+2A)^1_{s,0,-1}$} & {\footnotesize
${{N^{k,0,\epsilon}_{12}}_{|_{\epsilon=-1\atop{k<0}}}}$}&
\\

\vspace{1mm}

&{\footnotesize  $(A_{1,1}+2A)^2_{k,0,1}$} & {\footnotesize
${{N^{k,0,\epsilon}_{12}}_{|_{\epsilon=1\atop{k<0}}}}$}&
\\

\vspace{1mm}

&{\footnotesize  $(A_{1,1}+2A)^2_{s,0,-1}$} & {\footnotesize
${{N^{k,0,\epsilon}_{12}}_{|_{\epsilon=-1\atop{k>0}}}}$}&
\\

\vspace{2mm}

&{\footnotesize  $(A_{1,1}+2A)^2_{0,\epsilon,0}$} & {\footnotesize
${{N^{0,\epsilon,0}_{12}}}$}&
\smallskip \\

%%%%%%%%%%%%%%%%%%%%%%%%%%%%%%%%%%%%%%%%%%%%%%%%%%%%%%%%%%%%%%%%%%%%%%%%%%%%
{\footnotesize $C^5_p$}& {\footnotesize   $I_{(1,2)}$} &
{\footnotesize $A_{12}$}&
\\

\vspace{1mm}

{\footnotesize $p\geq 0$}&{\footnotesize
$(A_{1,1}+2A)^0_{0,0,\epsilon}$} & {\footnotesize
${{N^{k,0,\epsilon}_{12}}_{|_{k=o}}}$}&
\\

\vspace{1mm}

&{\footnotesize ${(A_{1,1}+2A)^1_{k,0,1}}_{|_{0<k \leq 1}}$} &
{\footnotesize ${{N^{k,0,\epsilon}_{12}}_{|_{\epsilon=1\atop{0<k
\leq 1}}}}$}&\\

\vspace{1mm}

&{\footnotesize ${(A_{1,1}+2A)^1_{k,0,1}}_{|_{k>1}}$} &
{\footnotesize Not exist}&\\

\vspace{1mm}

&{\footnotesize ${(A_{1,1}+2A)^1_{s,0,-1}}_{|_{-1<s<0}}$} &
{\footnotesize ${{N^{k,0,\epsilon}_{12}}_{|_{\epsilon=-1\atop{-1<k
<0}}}}$}&\\

\vspace{1mm}

&{\footnotesize ${(A_{1,1}+2A)^1_{s,0,-1}}_{|_{s<-1}}$} &
{\footnotesize Not exist}&\\

\vspace{1mm}

&{\footnotesize ${(A_{1,1}+2A)^1_{-1,0,-1}}$} & {\footnotesize
${{N^{-1,0,-1}_{12}}}$}&\\

\vspace{1mm}

&{\footnotesize ${(A_{1,1}+2A)^2_{k,0,1}}_{|_{-1<k<0}}$} &
{\footnotesize ${{N^{k,0,\epsilon}_{12}}_{|_{\epsilon=1\atop{-1<k
<0}}}}$}&\\

\vspace{1mm}

&{\footnotesize ${(A_{1,1}+2A)^2_{k,0,1}}_{|_{k \leq -1}}$} &
{\footnotesize Not exist}&\\

\vspace{1mm}

&{\footnotesize ${(A_{1,1}+2A)^2_{s,0,-1}}_{|_{0<s \leq 1}}$} &
{\footnotesize ${{N^{k,0,\epsilon}_{12}}_{|_{\epsilon=-1\atop{0<k
\leq 1}}}}$}&\\

\vspace{-1mm}

&{\footnotesize ${(A_{1,1}+2A)^2_{s,0,-1}}_{|_{s >1}}$} &
{\footnotesize Not exist}&\smallskip \\\hline\hline

\end{tabular}
\end{center}
Note that in Ref. \cite{H} $C^5_{p=0}$ and $C^2_{p=-1}$ appear
because of incorrect automorphism supergroups.
%%%%%%%%%%%%%%%%%%%%%%%%%%%%%%%%%%%%%%%%%%%%%%%%%%%%%%%%%%%%%%%%%%%%%%%%%%%%%%%%%%%%%%
%%%%%%%%%%%%%%%%%%%%%%%%%%%%%%%%%%%%%%%%%%%%%%%%%%%%%%%%%%%%%%%%%%%%%%%%%%%%%%%%%%%%%%
%%%%%%%%%%%%%%%%%%%%%%%%%%%%%%%%%%%%%%%%%%%%%%%%%%%%%%%%%%%%%%%%%%%%%%%%%%%%%%%%%%%%%
\newpage

{\bf Appendix D: Four and six dimensional Drinfel'd superdoubles (anti)commutation relations}\\

\vspace{5mm}

1.~(Anti)Commutation relations for Lie superalgebras ${\cal{D}}$
of the type $(2,2)$

\vspace{3mm}

\begin{tabular}{l l l p{2mm} }

\vspace{3mm}

${\cal{D}}=B\oplus I_{(1,1)}:$&&&\\

$\;\;\;\;\;\;\;[T_1 , T_3]=T_3,$&$\;\;[T_1 , T_4]=T_4,$&$\;\;\{T_3 ,T_4 \}=-iT_2.$&\\

\end {tabular}

\vspace{4mm}

\begin{tabular}{l l l p{2mm} }

\vspace{3mm}

${\cal{D}}=(A_{1,1}+A)\oplus I_{(1,1)}:$&&&\\

$\;\;\;\;\;\;\;[T_3 , T_2]=-T_4,$&$\;\;\;\;\{T_3 ,T_3 \}=iT_1.$&\\

\end {tabular}

\vspace{4mm}

\begin{tabular}{l l l lp{2mm} }

\vspace{3mm}

${\cal{D}}=B
\oplus (A_{1,1}+A):$& & & &\\

$\;\;\;\;\;\;[T_1 , T_3]=T_3,$&$\;[T_1 , T_4]=-T_3-T_4,$&$\;\;
\{T_3 ,T_4 \}=-iT_2,$&$\;\;\{T_4 ,T_4 \}=iT_2.$&\\

\end {tabular}

\vspace{4mm}

\begin{tabular}{l l l lp{2mm} }

\vspace{3mm}

${\cal{D}}=B
\oplus (A_{1,1}+A).i:$& & & &\\

$\;\;\;\;\;\;[T_1 , T_3]=T_3,$&$\;[T_1 , T_4]=T_3-T_4,$&$\;\;
\{T_3 ,T_4 \}=-iT_2,$&$\;\;\{T_4 ,T_4 \}=-iT_2.$&\\

\end {tabular}

\vspace{9mm}

2.~(Anti)Commutation relations for Lie superalgebras ${\cal{D}}$
of the type $(4,2)$

\vspace{5mm}

\begin{tabular}{l l l p{2mm} }

\vspace{2mm}

${\cal{D}}=(2A_{1,1}+A)\oplus I_{(2,1)}:$&&&\\

$\;\;\;\;\;\;\;[T_5 , T_3]=-T_6,$&$\{T_5 , T_5\}=iT_1.$&&\\

\end {tabular}

\vspace{4mm}

\begin{tabular}{l l l p{2mm} }

\vspace{2mm}

${\cal{D}}=(B+A_{1,1})\oplus I_{(2,1)}:$&&&\\

$\;\;\;\;\;\;\;[T_1 , T_5]=T_5,$&$\;\;\;\;\;[T_1 , T_6]=-T_6,$&$\{T_5 , T_6\}=-iT_3.$&\\

\end {tabular}

\vspace{4mm}

\begin{tabular}{l l l p{2mm} }

\vspace{2mm}

${\cal{D}}=C^1_p\oplus I_{(2,1)}:\;\;\;\;p\in \Re$&&&\\

\vspace{2mm}

$\;\;\;\;\;\;\;[T_1 , T_2]=T_2,$&$\;[T_1 , T_4]=-T_4,$&$\;\;[T_1 , T_5]=pT_5,$&\\
\vspace{1mm}
$\;\;\;\;\;\;\;[T_1 , T_6]=-pT_6,$&$\;[T_2 , T_4]=T_3,$&$\;\;\{T_5 ,T_6 \}=-ip T_3.$&\\

\end {tabular}

\vspace{4mm}

\begin{tabular}{l l l l p{2mm} }

\vspace{2mm}

${\cal{D}}=C^1_{\frac{1}{2}}\oplus I_{(2,1)}:$&&&\\

\vspace{2mm}

$\;\;\;\;\;\;\;[T_1 , T_2]=T_2,$&$\;[T_1 , T_4]=-T_4,$&$\;\;[T_1 ,
T_5]=\frac{1}{2} T_5,$&
$\;\;[T_1 , T_6]=-\frac{1}{2} T_6,$\\
\vspace{1mm}

$\;\;\;\;\;\;\;[T_2 , T_4]=T_3,$&$\;[T_5 , T_4]=-T_6,$&$\;\;\{T_5
,T_5 \}=i T_2,$&
$\;\;\{T_5 ,T_6 \}=-\frac{i}{2} T_3.$\\

\end {tabular}

\vspace{4mm}

\begin{tabular}{l l l  p{2mm} }

\vspace{2mm}

${\cal{D}}=(B+A_{1,1})\oplus (B+A_{1,1}).i:$&&&\\

\vspace{2mm}

$\;\;\;\;\;\;\;[T_1 , T_5]=T_5,$&$[T_1 , T_6]=-T_6,$&$\;\;[T_4 ,
T_6]= T_6,$&\\
\vspace{1mm}

$\;\;\;\;\;\;\;[T_5 , T_4]=T_5,$&$\{T_5
,T_6 \}=i T_2-iT_3.$&\\

\end {tabular}

\vspace{4mm}

\begin{tabular}{l l l l p{2mm} }

\vspace{2mm}

${\cal{D}}=(B+A_{1,1})\oplus (2A_{1,1}+A):$&&&\\

$\;\;\;\;\;\;\;[T_1 , T_5]=T_5,$&$[T_1 , T_6]=-T_5-T_6,$&$\;\{T_5
, T_6\}= -i T_3,$& $\{T_6
,T_6 \}=iT_3.$&\\

\end {tabular}

\vspace{4mm}

\begin{tabular}{l l l l p{2mm} }

\vspace{2mm}

${\cal{D}}=(B+A_{1,1})\oplus (2A_{1,1}+A).i:$&&&&\\

$\;\;\;\;\;\;\;[T_1 , T_5]=T_5,$&$[T_1 , T_6]=T_5-T_6,$&$\;\{T_5 ,
T_6\}= -i T_3,$& $\{T_6
,T_6 \}=-iT_3.$&\\

\end {tabular}

\vspace{4mm}

\begin{tabular}{l l l  p{2mm} }

\vspace{2mm}

${\cal{D}}=C^1_0\oplus C^1_{0,k}:\;\;\;\;k \in \Re-\{0\}$&&&\\

\vspace{2mm}

$\;\;\;\;\;\;\;[T_1 , T_2]=T_2,$&$[T_1 , T_4]=-T_4,$&$\;\;[T_2 ,
T_3]= k T_2,$&\\
\vspace{1mm}

$\;\;\;\;\;\;\;[T_3 , T_4]=k T_4,$&$[T_2
,T_4 ]=-k T_1+T_3.$&\\

\end {tabular}

\vspace{4mm}

\begin{tabular}{l l l l p{2mm} }

\vspace{2mm}

${\cal{D}}=C^1_p\oplus (2A_{1,1}+A):\;\;\;\;p\in \Re$&&&\\

\vspace{2mm}

$\;\;\;\;\;\;\;[T_1 , T_2]=T_2,$&$[T_1 , T_4]=-T_4,$&$\;\;[T_1 ,
T_5]= p T_5,$&$\;\;[T_2 ,
T_4]=  T_3,$&\\
\vspace{1mm}

$\;\;\;\;\;\;\;[T_1, T_6]=-T_5-p T_6,$&$\{T_5 , T_6\}= -ip
T_3,$&$\;\;\{T_6 ,
T_6\}= i T_3.$&\\

\end {tabular}

\vspace{4mm}

\begin{tabular}{l l l l p{2mm} }

\vspace{2mm}

${\cal{D}}=C^1_p\oplus (2A_{1,1}+A).i:\;\;\;\;p\in \Re$&&&\\

\vspace{2mm}

$\;\;\;\;\;\;\;[T_1 , T_2]=T_2,$&$[T_1 , T_4]=-T_4,$&$\;\;[T_1 ,
T_5]= p T_5,$&$\;\;[T_2 ,
T_4]=  T_3,$&\\
\vspace{1mm}

$\;\;\;\;\;\;\;[T_1, T_6]=T_5-p T_6,$&$\{T_5 , T_6\}= -ip
T_3,$&$\;\;\{T_6 ,
T_6\}= -i T_3.$&\\

\end {tabular}

\vspace{4mm}

\begin{tabular}{l l l l p{2mm} }

\vspace{2mm}

${\cal{D}}=C^1_{p=\frac{1}{2}}\oplus (2A_{1,1}+A).ii:$&&&\\

\vspace{2mm}

$\;\;\;\;\;\;\;[T_1 , T_2]=T_2,$&$[T_1 , T_4]=-T_4,$&$\;\;[T_1 ,
T_5]= \frac{1}{2} T_5,$&$\;\;[T_1 ,
T_6]= -\frac{1}{2} T_6,$&\\
\vspace{1mm}

$\;\;\;\;\;\;\;[T_2, T_4]=T_3,$&$[T_2, T_6]=-T_5,$&$\;\;\{T_5 ,
T_6\}= -\frac{i}{2} T_3,$&$\;\;\{T_6 ,
T_6\}= i T_4.$&\\

\end {tabular}

\vspace{4mm}

\begin{tabular}{l l l l p{2mm} }

\vspace{2mm}

${\cal{D}}=C^1_{p}\oplus C^1_{-p}.i:\;\;\;\;p\in \Re$&&&\\

\vspace{2mm}

$\;\;\;\;\;\;\;[T_1 , T_2]=T_2,$&$[T_1 , T_3]=T_2,$&$\;\;[T_1 ,
T_4]= -T_1-T_4,$&$\;\;[T_1 ,
T_5]= p T_5,$&\\
\vspace{2mm}

$\;\;\;\;\;\;\;[T_1, T_6]=-pT_6,$&$[T_2, T_4]=T_3,$&$\;\;[T_3 ,
T_4]= T_3,$&$\;\;[T_4 ,
T_6]= p T_6,$&\\
\vspace{1mm}

$\;\;\;\;\;\;\;[T_5, T_4]=pT_5,$&$\{T_5 , T_6\}= ipT_2-ipT_3.$&&&\\

\end {tabular}

\vspace{4mm}

\begin{tabular}{l l l l p{2mm} }

\vspace{2mm}

${\cal{D}}=C^1_{\frac{1}{2}}\oplus C^1_{p=-\frac{1}{2}}.i:$&&&\\

\vspace{2mm}

$\;\;\;\;\;\;\;[T_1 , T_2]=T_2,$&$[T_1 , T_3]=T_2,$&$\;\;[T_1 ,
T_4]= -T_1-T_4,$&$\;\;[T_1 ,
T_5]= \frac{1}{2} T_5,$&\\
\vspace{2mm}

$\;\;\;\;\;\;\;[T_1, T_6]=-\frac{1}{2}T_6,$&$[T_2,
T_4]=T_3,$&$\;\;[T_3 , T_4]= T_3,$&$\;\;[T_4 ,
T_6]= \frac{1}{2} T_6,$&\\
\vspace{1mm}

$\;\;\;\;\;\;\;[T_5, T_4]=\frac{1}{2}T_5-T_6,$&$\{T_5 , T_5\}=
iT_2,$&$\;\;\{T_5 , T_6\}= \frac{i}{2}T_2-
\frac{i}{2}T_3.$&&\\

\end {tabular}

\vspace{4mm}

\begin{tabular}{l l l l p{2mm} }

\vspace{2mm}

${\cal{D}}=C^1_{\frac{1}{2}}\oplus C^1_{p=-\frac{1}{2}}.ii:$&&&\\

\vspace{2mm}

$\;\;\;\;\;\;\;[T_1 , T_2]=T_2,$&$[T_1 , T_3]=-T_2,$&$\;\;[T_1 ,
T_4]= T_1-T_4,$&$\;\;[T_1 ,
T_5]= \frac{1}{2} T_5,$&\\
\vspace{2mm}

$\;\;\;\;\;\;\;[T_1, T_6]=-\frac{1}{2}T_6,$&$[T_2,
T_4]=T_3,$&$\;\;[T_3 , T_4]= -T_3,$&$\;\;[T_4 ,
T_6]= -\frac{1}{2} T_6,$&\\
\vspace{1mm}

$\;\;\;\;\;\;\;[T_5, T_4]=-\frac{1}{2}T_5-T_6,$&$\{T_5 , T_5\}=
iT_2,$&$\;\;\{T_5 , T_6\}= -\frac{i}{2}T_2-
\frac{i}{2}T_3.$&&\\

\end {tabular}

\vspace{4mm}

\begin{tabular}{l l l l p{2mm} }

\vspace{2mm}

${\cal{D}}=C^1_{\frac{1}{2}}\oplus C^1_{\frac{1}{2}}.i:$&&&\\

\vspace{2mm}

$\;\;\;\;\;\;\;[T_1 , T_2]=T_2,$&$[T_1 , T_3]=T_2,$&$\;\;[T_1 ,
T_4]= -T_1-T_4,$&$\;\;[T_1 ,
T_5]= \frac{1}{2} T_5,$&\\
\vspace{2mm}

$\;\;\;\;\;\;\;[T_1, T_6]=-T_5-\frac{1}{2}T_6,$&$[T_2,
T_4]=T_3,$&$\;\;[T_3 , T_4]= T_3,$&$\;\;[T_4 ,
T_6]= -\frac{1}{2} T_6,$&\\
\vspace{1mm}

$\;\;\;\;\;\;\;[T_5, T_4]=-\frac{1}{2}T_5-T_6,$&$\{T_5 , T_5\}=
iT_2,$&$\;\;\{T_5 , T_6\}= -\frac{i}{2}T_2-
\frac{i}{2}T_3,$&$\;\;\{T_6 , T_6\}=i T_3.$&\\

\end {tabular}

\vspace{3mm}

\begin{tabular}{l l l l p{2mm} }

\vspace{2mm}

${\cal{D}}=C^1_{\frac{1}{2}}\oplus C^1_{\frac{1}{2}}.ii:$&&&\\

\vspace{2mm}

$\;\;\;\;\;\;\;[T_1 , T_2]=T_2,$&$[T_1 , T_3]=-T_2,$&$\;\;[T_1 ,
T_4]= T_1-T_4,$&$\;\;[T_1 ,
T_5]= \frac{1}{2} T_5,$&\\
\vspace{2mm}

$\;\;\;\;\;\;\;[T_1, T_6]= T_5-\frac{1}{2}T_6,$&$[T_2,
T_4]=T_3,$&$\;\;[T_3 , T_4]= -T_3,$&$\;\;[T_4 ,
T_6]= \frac{1}{2} T_6,$&\\
\vspace{1mm}

$\;\;\;\;\;\;\;[T_5, T_4]= \frac{1}{2}T_5-T_6,$&$\{T_5 , T_5\}=
iT_2,$&$\;\;\{T_5 , T_6\}= \frac{i}{2}T_2-
\frac{i}{2}T_3,$&$\;\;\{T_6 , T_6\}=-i T_3.$&\\

\end {tabular}

\vspace{3mm}

\begin{tabular}{l l l l p{2mm} }

\vspace{2mm}

${\cal{D}}=C^1_{\frac{1}{2}}\oplus C^1_{\frac{1}{2},k}:\;\;\;\;k \in \Re-\{0\}$&&&\\

\vspace{2mm}

$\;\;\;\;\;\;\;[T_1 , T_2]=T_2,$&$[T_1 , T_4]=-T_4,$&$\;\;[T_1 ,
T_5]= \frac{1}{2} T_5,$&$\;\;[T_1, T_6]= -\frac{1}{2}T_6,$&\\
\vspace{2mm}

$\;\;\;\;\;\;\;[T_2, T_6]= -kT_5,$&$[T_2,
T_4]=-kT_1+T_3,$&$\;\;[T_3 , T_4]= kT_4,$&$\;\;[T_3 ,
T_6]= \frac{k}{2} T_6,$&\\
\vspace{2mm}

$\;\;\;\;\;\;\;[T_2, T_3]= kT_2,$&$[T_5 , T_3]= \frac{k}{2}
T_5,$&$\;\;[T_5 , T_4]=-T_6,$&$\;\;\{T_5 , T_5\}=i T_2,$&\\
\vspace{1mm}

$\;\;\;\;\;\;\;\{T_5 , T_6\}= \frac{ik}{2}T_1-
\frac{i}{2}T_3,$&$\{T_6 , T_6\}= ikT_4.$&&&\\

\end {tabular}
%%%%%%%%%%%%%%%%%%%%%%%%%%%%%%%%%%%%%%%%%%%%%%%%%%%%%%%%%%%%%%%
\vspace{3mm}

3.~(Anti)Commutation relations for Lie superalgebras ${\cal{D}}$
of the type $(2,4)$. Note that ${\tilde {\cal
G}}_{\alpha\beta\gamma}$ is one of the dual Lie superalgebras
$(A_{1,1}+2A)^0_{\alpha ,\beta,\gamma }\;,\;(A_{1,1}+2A)^1_{\alpha
,\beta,\gamma }$ or $(A_{1,1}+2A)^2_{\alpha ,\beta,\gamma }.$

\vspace{2mm}

\begin{tabular}{l l l l p{2mm} }

\vspace{2mm}

${\cal{D}}=C^2_{p}\oplus {\tilde {\cal G}}_{\alpha, \beta, \gamma}:\;\;\;\;p \in [-1 , 1]$&&&\\

\vspace{2mm}

$\;\;\;\;\;\;\;[T_1 , T_3]=T_3,$&$[T_1 , T_4]=pT_4,$&$\;\;\{T_4 ,
T_6\}= -ip T_2,$&$\;\;[T_1, T_5]= -\alpha T_3-\beta T_4 -T_5,$&\\
\vspace{2mm}

$\;\;\;\;\;\;\;\{T_3, T_5\}= -iT_2,$&$\{T_5, T_5\}= i\alpha
T_2,$&$\;\;\{T_5, T_6 \}= i\beta T_2,$&$\;\;[T_1 ,
T_6]= -\gamma T_4 -\beta T_3 -pT_6,$&\\
\vspace{2mm}

$\;\;\;\;\;\;\;\{T_6, T_6 \}= i\gamma T_2.$&&&&\\

\end {tabular}

\vspace{4mm}

\begin{tabular}{l l l l p{2mm} }

\vspace{2mm}

${\cal{D}}=C^3\oplus {\tilde {\cal G}}_{\alpha, \beta, \gamma}:$&&&\\

\vspace{2mm}

$\;\;\;\;\;\;\;[T_1 , T_4]=T_3,$&$\{T_4, T_5\}=
-iT_2,$&$\;\;\{T_5 ,
T_5\}= i \alpha T_2,$&$\;\;[T_1, T_5]= -\alpha T_3-\beta T_4 -T_6,$&\\
\vspace{2mm}

$\;\;\;\;\;\;\;\{T_5, T_6\}= i \beta T_2,$&$\{T_6, T_6\}= i\gamma
T_2,$&$\;\;[T_1 , T_6]= -\beta T_3 -\gamma T_4.$&&\\

\end {tabular}

\vspace{4mm}

\begin{tabular}{l l l l p{2mm} }

\vspace{2mm}

${\cal{D}}=C^4\oplus {\tilde {\cal G}}_{\alpha, \beta, \gamma}:$&&&\\

\vspace{2mm}

$\;\;\;\;\;\;\;[T_1 , T_4]=T_3+T_4,$&$[T_1 ,
T_3]=T_3,$&$\;\;\{T_3 ,
T_5\}= -i T_2,$&$\;\;[T_1, T_5]= -\alpha T_3-\beta T_4 -T_5-T_6,$&\\
\vspace{2mm}

$\;\;\;\;\;\;\;\{T_4, T_5\}= -iT_2,$&$\{T_4,
T_6\}=-iT_2,$&$\;\;\{T_5, T_5 \}= i\alpha T_2,$&$\;\;[T_1 ,
T_6]= -\gamma T_4 -\beta T_3 -T_6,$&\\
\vspace{2mm}

$\;\;\;\;\;\;\;\{T_5, T_6 \}= i\beta T_2,$&$\{T_6 , T_6 \}= i\gamma T_2.$ &&&\\

\end {tabular}

\vspace{4mm}

\begin{tabular}{l l l l p{2mm} }

\vspace{2mm}

${\cal{D}}=C^5_{p}\oplus {\tilde {\cal G}}_{\alpha, \beta, \gamma}:\;\;\;\;p\geq 0$&&&\\

\vspace{2mm}

$\;\;\;\;\;\;\;[T_1 , T_3]=pT_3-T_4,$&$\{T_3, T_5
\}=-ipT_2,$&$\;\;\{T_3 ,
T_6\}= i T_2,$&$\;\;[T_1, T_5]= -\alpha T_3-\beta T_4 -pT_5-T_6,$&\\
\vspace{2mm}

$\;\;\;\;\;\;\;[T_1 , T_4]=T_3+pT_4,$&$\{T_4, T_5\}=-i
T_2,$&$\;\;\{T_4, T_6 \}= -ipT_2,$&$\;\;[T_1 ,
T_6]= -\gamma T_4 -\beta T_3 -pT_6+T_5,$&\\
\vspace{2mm}

$\;\;\;\;\;\;\;\{T_5, T_5 \}= i\alpha T_2,$&$\{T_5, T_6 \}=
i\beta T_2,$&$\;\;\{T_6, T_6 \}= i\gamma T_2.$&&\\

\end {tabular}

\vspace{4mm}

\begin{tabular}{l l l p{2mm} }

\vspace{2mm}

${\cal{D}}={(A_{1,1}+2A)}^0 \oplus I_{(1,2)}:$&\\

\vspace{2mm}

$\;\;\;\;\;\;\;[T_3 , T_2]=-T_5,$&$\{T_3, T_3 \}=iT_1.$&\\

\end {tabular}

\vspace{4mm}

\begin{tabular}{l l l l p{2mm} }

\vspace{2mm}

${\cal{D}}={(A_{1,1}+2A)}^1 \oplus I_{(1,2)}:$& &&\\

\vspace{2mm}

$\;\;\;\;\;\;\;[T_3 , T_2]=-T_5,$&$[T_4 , T_2]=-T_6,$&$\;\;\{T_3,
T_3 \} = iT_1,$ &$\;\;\{T_4, T_4 \}=iT_1.$ &\\

\end {tabular}

\vspace{4mm}

\begin{tabular}{l l l l p{2mm} }

\vspace{2mm}

${\cal{D}}={(A_{1,1}+2A)}^2 \oplus I_{(1,2)}:$& &&\\

\vspace{2mm}

$\;\;\;\;\;\;\;[T_3 , T_2]=-T_5,$&$[T_4 , T_2]=T_6,$&$\;\;\{T_3,
T_3 \} = iT_1,$ &$\;\;\{T_4, T_4 \}=-iT_1.$ &\\

\end {tabular}

%%%%%%%%%%%%%%%%%%%%%%%%%%%%%%%%%%%%%%%%%%%%%%%%%%%%%%%%%%%%%%%%%%%%%%%%%%%%%%%%%%%%%%%
\newpage

{\bf Appendix E: Comparison of theorem 3 with those of Ref. \cite
{H}}

\vspace{5mm}

1.~Comparison of $\cal{D}$$sd^{3\;\;p}_{(2,4)}$  with
$DD_{(1,2)}II_{p}:$

\smallskip

In $DD_{(1,2)}II_{p}$ the Drinfel'd superdoubles $\Big( C^2_p ,
(A_{1,1} +2A)^0_{-1,-1,-1}\Big) \cong \Big( C^2_p |
{N^{-1,-1,-1}_{12}}\Big)$,\;$\Big( C^2_p , {(A_{1,1}
+2A)^1_{\epsilon,k,\epsilon}}_{|_{-1<k<0}}\Big)\\ \cong \Big(
C^2_p | {N^{\epsilon,k,\epsilon}_{12}}_{|_{-1<k<0}}\Big)$ and
$\Big( C^2_p , {(A_{1,1}
+2A)^2_{\epsilon,s,-\epsilon}}_{|_{s<0}}\Big)$ do not exist. On
the other hand, in $\cal{D}$$sd^{3\;\;p}_{(2,4)}$ the Drinfel'd
superdouble $\Big( C^2_p | {{N^{\epsilon,k,\epsilon}_{12}}}_{|_{k
\geq 1}}\Big) \cong \Big( C^2_p , {(A_{1,1}
+2A)^1_{\epsilon,k,\epsilon}}_{|_{{k \geq 1}}}\Big)$ does not
exist.

\vspace{4mm}

2.~Comparison of $\cal{D}$$sd^{3\;\;p=1}_{(2,4)}$  with
$DD_{(1,2)}II_{1}:$

\smallskip

In $DD_{(1,2)}II_{1}$ the Drinfel'd superdoubles $\Big( C^2_1 ,
(A_{1,1} +2A)^2_{-1,0,1}\Big)$,\;$\Big( C^2_ {p=-1 } , (A_{1,1}
+2A)^0_{\epsilon,0,0}\Big)$ and $\Big( C^2_ {p=-1 } , (A_{1,1}
+2A)^1_{-1,0,-1}\Big)$ do not exist.

\vspace{3mm}

3.~There are not differences between
$\cal{D}$$sd^{3\;\;p=0}_{(2,4)}$ with $DD_{(1,2)}II_{0}.$\\

4.~There are not differences between $\cal{D}$$sd^{4}_{(2,4)}$
with $DD_{(1,2)}III$.\\

5.~Comparison of $\cal{D}$$sd^{6}_{(2,4)}$  with $DD_{(1,2)}IV:$

\smallskip

In $DD_{(1,2)}IV$ the Drinfel'd superdoubles $\Big( C^2_{p=-1} ,
(A_{1,1} +2A)^0_{-1,-1,-1}\Big)$,\;$\Big( C^2_ {p=-1 } , (A_{1,1}
+2A)^2_{\epsilon,1,0}\Big)$,\; $\Big( C^2_ {p=-1 } , \\{(A_{1,1}
+2A)^2_{\epsilon,k,\epsilon}}_{|_{-1<k<0}}\Big)$,\;$\Big( C^2_
{p=-1 } , {(A_{1,1}+2A)^1_{-1,k,-1}}_{|_{0<k<1}}\Big)$ and $\Big(
C^2_ {p=-1 } , {(A_{1,1}
+2A)^2_{\epsilon,s,-\epsilon}}_{|_{s<0}}\Big)$ do not exist. On
the other hand, in $\cal{D}$$sd^{6}_{(2,4)}$ the Drinfel'd
superdouble $\Big( C^2_{p=-1} | {{N^{1,k,1}_{12}}}_{|_{k \geq
1}}\Big) \cong \Big( C^2_{p=-1} , {(A_{1,1}
+2A)^1_{1,k,1}}_{|_{{k \geq 1}}}\Big)$ does not exist.\\

6.~Comparison of $\cal{D}$$sd^{7\;\;p}_{(2,4)}$  with
$DD_{(1,2)}V_{p}:$

\smallskip

In $DD_{(1,2)}V_{p}$ the Drinfel'd superdoubles $\Big( C^5_{p} ,
{(A_{1,1} +2A)^1_{k,0,1}}_{|_{k>1}}\Big)$,\;$\Big( C^5_{p} ,
{(A_{1,1} +2A)^1_{k,0,-1}}_{|_{k<-1}}\Big)$,\; $\Big( C^5_{p} \;,\\
{(A_{1,1} +2A)^2_{s,0,1}}_{|_{s\leq -1}}\Big)$ and $\Big( C^5_{p}
, {(A_{1,1} +2A)^2_{s,0,-1}}_{|_{s>1}}\Big)$  do not exist.\\

7.~Comparison of $\cal{D}$$sd^{8}_{(2,4)}$  with
$DD_{(1,2)}V_{0}:$

\smallskip

In $DD_{(1,2)}V_{0}$ the Drinfel'd superdoubles $\Big( C^5_{0} ,
{(A_{1,1} +2A)^2_{1,0,-1}}\Big)$ do not exist.\\

8.~Comparison of $\cal{D}$$sd^{2}_{(2,4)}$  with $DD_{(1,2)}VI:$

\smallskip

In $DD_{(1,2)}VI$ the Drinfel'd superdoubles $\Big( C^2_{0} ,
{(A_{1,1} +2A)^0_{-1,-1,-1}}\Big)$,\;$\Big( C^2_{0} , {(A_{1,1}
+2A)^1_{\epsilon,k,\epsilon}}_{|_{-1<k<0}}\Big)$ and  $\Big(
C^2_{0}\; , \\{(A_{1,1}
+2A)^1_{\epsilon,s,-\epsilon}}_{|_{s<0}}\Big)\;\;$ do not exist.
On the other hand, in $\cal{D}$$sd^{2}_{(2,4)}$ the Drinfel'd
superdouble $\Big( C^2_{0}\; , \\{(A_{1,1}
+2A)^1_{\epsilon,k,\epsilon}}_{|_{k\geq 1}}\Big)$ does not exist.\\

9.~Comparison of $\cal{D}$$sd^{5}_{(2,4)}$  with $DD_{(1,2)}VII:$

\smallskip

In $DD_{(1,2)}VII$ the Drinfel'd superdoubles $\Big( C^3 ,
{(A_{1,1} +2A)^1_{-1,0,-1}}\Big)$ and $\Big( C^3 , {(A_{1,1}
+2A)^2_{1 ,0,-1}}\Big)$ do not exist.\\

10.~Comparison of $\cal{D}$$sd^{7\;\;p=0}_{(2,4)}$  with
$DD_{(1,2)}VIII:$

\smallskip

In $DD_{(1,2)}VIII$ the Drinfel'd superdoubles $\Big( C^5_{0} ,
{(A_{1,1} +2A)^1_{k,0,1}}_{|_{k>1}}\Big)$,\;\;$\Big( C^5_{0} ,
{(A_{1,1} +2A)^0_{0,0,-1}}\Big)$,\;\;$\Big( C^5_{0}\; ,\\
{(A_{1,1} +2A)^1_{k,0,-1}}_{|_{k<0}}\Big)$,\; $\Big( C^5_{0} ,
{(A_{1,1} +2A)^2_{s,0,1}}_{|_{s<-1}}\Big)$ and $\Big( C^5_{0} ,
{(A_{1,1}
+2A)^2_{s,0,-1}}_{|_{s>0}}\Big)$ do not exist.\\
%%%%%%%%%%%%%%%%%%%%%%%%%%%%%%%%%%%%%%%%%%%%%%%%%%%%%%%%%%%%%%%%%%%%%%%%%%%%%%%%%%%%%%%5
\bigskip
\vspace{9mm}

{\bf Appendix F: Isomorphisms of the Drinfeld's superdoubles of
the types $(2,2), \;(4,2)$ and $(2,4)$} \vspace{5mm}

\smallskip
Isomorphisms of the Drinfeld's superdoubles of the type $(2,2)$\\

{\bf $\cal{D}$$sd^{3}_{(2,2)}:$}\\

{\footnotesize $\Big(B, I_{(1,1)}\Big)\longrightarrow
\Big(B,(A_{1,1}+A)\Big) $}

$$
{\footnotesize C=\left( \begin{tabular}{cccc}
              $ 1$ & $c$& $ 0$& $ 0$ \\
              $0$ & $ab$& $ 0$& $ 0$ \\
               $0$ & $0$& $a$& $ 0$ \\
                 $0$ & $0$& $-\frac{a}{2}$& $ b$ \\
                 \end{tabular} \right),\;\;\;\;a,
                 b\in\Re-\{0\};\;\;
                 c\in\Re,}
$$

\vspace{0.60cm}

{\footnotesize $\Big(B, I_{(1,1)}\Big)\longrightarrow
\Big(B,(A_{1,1}+A).i\Big) $}

$$
{\footnotesize C=\left( \begin{tabular}{cccc}
              $ 1$ & $c$& $ 0$& $ 0$ \\
              $0$ & $ab$& $ 0$& $ 0$ \\
               $0$ & $0$& $a$& $ 0$ \\
                 $0$ & $0$& $\frac{a}{2}$& $ b$ \\
                 \end{tabular} \right),\;\;\;\;a,
                 b\in\Re-\{0\};\;\;
                 c\in\Re.}
$$

%%%%%%%%%%%%%%%%%%%%%%%%%%%%%%%%%%%%%%%%%%%%%%%%%%%%%%%%%%%%%%%%%%%%%%%%%%%%%%%%%%%%%
\vspace{0.90cm}

Isomorphisms of the Drinfeld's superdoubles of the type $(4,2)$\\

\vspace{0.50cm}

{\bf $\cal{D}$$sd^{3}_{(4,2)}:$}\\

{\footnotesize $\Big((B+A_{1,1}) , I_{(2,1)}\Big)\longrightarrow
\Big((B+A_{1,1}) , (B+A_{1,1}).i\Big) $}

$$
{\footnotesize C=\left( \begin{tabular}{cccccc}
              $ 1$ & $e$& $f$& $g$ & $ 0$& $ 0$\\
              $0$ & $d$& $a-bc$& $n$& $ 0$& $ 0$ \\
               $0$ & $d$& $a$& $n$& $0$& $ 0$ \\
                 $-1$ & $m$& $r$& $s$& $ 0$& $ 0$ \\
                 $0$ & $0$& $ 0$& $0$& $b$& $ 0$ \\
                 $0$ & $0$& $ 0$& $ 0$& $0$& $c$ \\
                 \end{tabular} \right),\;\;\;mn- ds\neq 0;\;\;\;b,
                 c\in\Re-\{0\};\;\;
                 a, d, e,...,s\in \Re,}
$$

\vspace{0.80cm}

{\footnotesize $\Big((B+A_{1,1}) , I_{(2,1)}\Big)\longrightarrow
\Big((B+A_{1,1}) , (2A_{1,1}+A)\Big) $}

$$
{\footnotesize C=\left( \begin{tabular}{cccccc}
              $1$ & $c$& $d$& $e$ & $ 0$& $ 0$\\
              $0$ & $m$& $r$& $v$& $ 0$& $ 0$ \\
              $0$ & $0$& $ab$ &$0$& $0$& $ 0$ \\
              $0$ & $u$& $s$& $n$& $ 0$& $ 0$ \\
              $0$ & $0$& $0$& $0$& $a$& $ 0$ \\
              $0$ & $0$& $0$& $ 0$& $-\frac{a}{2}$& $b$ \\
                 \end{tabular} \right),\;\;\;\;mn - uv\neq 0;\;\;\;a, b\in\Re-\{0\}; \;\;\;
                 c, d, e,...,v \in \Re,}
$$

\vspace{0.80cm}

{\footnotesize $\Big((B+A_{1,1}) , I_{(2,1)}\Big)\longrightarrow
\Big((B+A_{1,1}) , (2A_{1,1}+A).i\Big) $}

$$
{\footnotesize C=\left( \begin{tabular}{cccccc}
              $1$ & $c$& $d$& $e$ & $ 0$& $ 0$\\
              $0$ & $m$& $r$& $v$& $ 0$& $ 0$ \\
              $0$ & $0$& $ab$ &$0$& $0$& $ 0$ \\
              $0$ & $u$& $s$& $n$& $ 0$& $ 0$ \\
              $0$ & $0$& $0$& $0$& $a$& $ 0$ \\
              $0$ & $0$& $0$& $ 0$& $\frac{a}{2}$& $b$ \\
                 \end{tabular} \right),\;\;\;\;mn - uv\neq 0;\;\;\;a, b\in\Re-\{0\}; \;\;\;
                 c, d, e,...,v \in \Re,}
$$

\newpage

{\bf $\cal{D}$$sd^{4\;\;p=0}_{(4,2)}:$}\\

{\footnotesize $\Big(C^1_{p=0} , I_{(2,1)}\Big) \longrightarrow
\Big(C^1_{p=0} , C^1_{-p=0}.i\Big) $}

$$
{\footnotesize C=\left( \begin{tabular}{cccccc}
              $1$ & $-a$& $a(b+c)-d$& $c$ & $ 0$& $ 0$\\
              $0$ & $e$& $-ec$& $0$& $ 0$& $ 0$ \\
               $0$ & $e$& $eb$& $0$& $0$& $ 0$ \\
               $-1$ & $a$& $d$& $ b$& $ 0$& $ 0$ \\
                 $0$ & $0$& $ 0$& $0$& $m$& $s$ \\
                 $0$ & $0$& $ 0$& $ 0$& $r$& $n$ \\
                 \end{tabular} \right),\;\;\;\;mn-rs \neq 0;\;\;\;b+c \neq 0;\;\;\;
                 e\in\Re-\{0\};\;\;a,...,s\in\Re,}
$$

\vspace{0.8cm}

{\bf $\cal{D}$$sd^{4\;\;p}_{(4,2)}:\;\;\;\;\;\;\;p\in\Re-\{0\},$}\\

{\footnotesize $\Big(C^1_p , I_{(2,1)}\Big) \longrightarrow
\Big(C^1_{-p} , I_{(2,1)}\Big) $}

$$
{\footnotesize C=\left( \begin{tabular}{cccccc}
              $1$ & $d$& $e$& $f$ & $ 0$& $ 0$\\
              $0$ & $-\frac{bc}{a}$& $\frac{bcf}{a}$& $0$& $ 0$& $ 0$ \\
              $0$ & $0$& $-bc$& $0$& $0$& $ 0$ \\
              $0$ & $0$& $-ad$& $a$& $ 0$& $ 0$ \\
              $0$ & $0$& $ 0$& $0$& $0$& $b$ \\
              $0$ & $0$& $ 0$& $ 0$& $c$& $0$ \\
                 \end{tabular}
                 \right),\;\;\;\;\;a,b,c\in\Re-\{0\};\;\;\;
                 d,e,f\in\Re,}
$$

\vspace{0.8cm}

{\footnotesize $\Big(C^1_p , I_{(2,1)}\Big) \longrightarrow
\Big(C^1_{p} , C^1_{-p}.i\Big) $}

$$
{\footnotesize C=\left( \begin{tabular}{cccccc}
              $1$ & $-e$& $\frac{bce}{d}-n$& $\frac{bc}{d}-a$ & $ 0$& $ 0$\\
              $0$ & $d$& $ad-bc$& $0$& $ 0$& $ 0$ \\
               $0$ & $d$& $ad$& $0$& $0$& $ 0$ \\
               $-1$ & $e$& $n$& $ a$& $ 0$& $ 0$ \\
                 $0$ & $0$& $ 0$& $0$& $b$& $0$ \\
                 $0$ & $0$& $ 0$& $ 0$& $0$& $c$ \\
                 \end{tabular} \right),\;\;\;\;b,c,d \in\Re-\{0\};\;\;
                 a,e,n\in\Re,}
$$

\vspace{0.8cm}

{\footnotesize $\Big(C^1_p , I_{(2,1)}\Big) \longrightarrow
\Big(C^1_{p} , (2A_{1,1}+A)\Big) $}

$$
{\footnotesize C=\left( \begin{tabular}{cccccc}
              $1$ & $d$& $e$& $f$ & $ 0$& $ 0$\\
              $0$ & $\frac{ab}{c}$& $-\frac{abf}{c}$& $0$& $ 0$& $ 0$ \\
               $0$ & $0$& $ab$& $0$& $0$& $0$ \\
               $0$ & $0$&$ -cd$& $ c$& $ 0$& $ 0$ \\
                 $0$ & $0$& $ 0$& $0$& $a$& $0$ \\
                 $0$ & $0$& $ 0$& $ 0$& $-\frac{a}{2p}$& $b$ \\
                 \end{tabular} \right),\;\;\;\;\;\;a,b,c\in\Re-\{0\};\;\;
                 d,e,f\in\Re,}
$$

\vspace{0.70cm}

{\footnotesize $\Big(C^1_p , I_{(2,1)}\Big) \longrightarrow
\Big(C^1_{p} , (2A_{1,1}+A).i\Big) $}

$$
{\footnotesize C=\left( \begin{tabular}{cccccc}
              $1$ & $d$& $e$& $f$ & $ 0$& $ 0$\\
              $0$ & $\frac{ab}{c}$& $-\frac{abf}{c}$& $0$& $ 0$& $ 0$ \\
               $0$ & $0$& $ab$& $0$& $0$& $0$ \\
               $0$ & $0$&$ -cd$& $ c$& $ 0$& $ 0$ \\
                 $0$ & $0$& $ 0$& $0$& $a$& $0$ \\
                 $0$ & $0$& $ 0$& $ 0$& $\frac{a}{2p}$& $b$ \\
                 \end{tabular} \right),\;\;\;\;\;\;a,b,c\in\Re-\{0\};\;\;
                 d,e,f\in\Re,}
$$
\newpage

{\bf $\cal{D}$$sd^{5}_{(4,2)}:$}\\

{\footnotesize $\Big(C^1_{p=0},(2A_{1,1}+A) \Big) \longrightarrow
\Big(C^1_{p=0},(2A_{1,1}+A).i\Big) $}

$$
{\footnotesize C=\left( \begin{tabular}{cccccc}
              $1$ & $c$& $e$& $d$ & $ 0$& $ 0$\\
              $0$ & $-\frac{a^2}{b}$& $\frac{a^2d}{b}$& $0$& $ 0$& $ 0$ \\
               $0$ & $0$& $-a^2$& $0$& $0$& $0$ \\
               $0$ & $0$& $-bc$& $b$& $ 0$& $ 0$ \\
                 $0$ & $0$& $ 0$& $0$& $-a$& $0$ \\
                 $0$ & $0$& $ 0$& $ 0$& $n$& $a$ \\
                 \end{tabular} \right),\;\;\;\;a,b\in\Re-\{0\}; \;\;
                 c,d,e,n\in\Re,}
$$

\vspace{0.50cm}

{\bf $\cal{D}$$sd^{7}_{(4,2)}:$}\\

{\footnotesize $\Big(C^1_{\frac{1}{2}},I_{(2,1)} \Big)
\longrightarrow \Big(C^1_{p={\frac{1}{2}}},(2A_{1,1}+A).ii \Big)
$}

$$
{\footnotesize C=\left( \begin{tabular}{cccccc}
              $-1$ & $a$& $e$& $c$ & $ 0$& $ 0$\\
              $0$ & $0$& $ad$& $d$& $ 0$& $ 0$ \\
               $0$ & $0$& $-db^2$& $0$& $0$& $0$ \\
               $0$ & $b^2$& $cb^2$& $0$& $ 0$& $ 0$ \\
                 $0$ & $0$& $ 0$& $0$& $0$& $-bd$ \\
                 $0$ & $0$& $ 0$& $ 0$& $b$& $-bc$ \\
                 \end{tabular} \right),\;\;\;\;b,d\in\Re-\{0\}; \;\;
                 a,c,e\in\Re,}
$$

\vspace{0.40cm}

{\footnotesize $\Big(C^1_{\frac{1}{2}},I_{(2,1)} \Big)
\longrightarrow
\Big(C^1_{\frac{1}{2}},C^1_{p=-{\frac{1}{2}}}.i\Big) $}

$$
{\footnotesize C=\left( \begin{tabular}{cccccc}
              $1$ & $a$& $d$& $c$ & $ 0$& $ 0$\\
              $0$ & $b^2$& $-b^2c$& $0$& $ 0$& $ 0$ \\
               $0$ & $b^2$& $b^2e$& $0$& $0$& $0$ \\
               $-1$ & $-a$& $-a(e+c)-d$& $e$& $ 0$& $ 0$ \\
                 $0$ & $0$& $ 0$& $0$& $b$& $bc$ \\
                 $0$ & $0$& $ 0$& $ 0$& $0$& $b(e+c)$ \\
                 \end{tabular} \right),\;\;\;\;e+c \neq
                 0;\;\;b\in\Re-\{0\};\;\;\;
                a,c,d,e\in\Re,}
$$

\vspace{0.40cm}

{\footnotesize $\Big(C^1_{\frac{1}{2}},I_{(2,1)} \Big)
\longrightarrow
\Big(C^1_{\frac{1}{2}},C^1_{p=-{\frac{1}{2}}}.ii\Big) $}

$$
{\footnotesize C=\left( \begin{tabular}{cccccc}
              $1$ & $a$& $d$& $c$ & $ 0$& $ 0$\\
              $0$ & $b^2$& $-b^2c$& $0$& $ 0$& $ 0$ \\
               $0$ & $-b^2$& $b^2e$& $0$& $0$& $0$ \\
               $1$ & $a$& $-a(e-c)+d$& $e$& $ 0$& $ 0$ \\
                 $0$ & $0$& $ 0$& $0$& $b$& $bc$ \\
                 $0$ & $0$& $ 0$& $ 0$& $0$& $b(e-c)$ \\
                 \end{tabular} \right),\;\;\;\;e-c \neq
                 0;\;\;b\in\Re-\{0\};\;\;\;
                a,c,d,e\in\Re,}
$$

\vspace{0.40cm}

{\footnotesize $\Big(C^1_{\frac{1}{2}},I_{(2,1)} \Big)
\longrightarrow \Big(C^1_{\frac{1}{2}},C^1_{\frac{1}{2}}.i\Big) $}

$$
{\footnotesize C=\left( \begin{tabular}{cccccc}
              $1$ & $a$& $e$& $b$ & $ 0$& $ 0$\\
              $0$ & $c^2$& $-bc^2$& $0$& $ 0$& $ 0$ \\
               $0$ & $c^2$& $c^2d$& $0$& $0$& $0$ \\
               $-1$ & $-a$& $-a(d+b)-e$& $ d$& $ 0$& $ 0$ \\
                 $0$ & $0$& $ 0$& $0$& $c$& $bc$ \\
                 $0$ & $0$& $ 0$& $ 0$& $-c$& $cd$ \\
                 \end{tabular} \right),\;\;\;\;b+d \neq 0;\;\;c\in\Re-\{0\};\;\;
                 a,b,d,e\in\Re,}
$$

\vspace{0.40cm}

{\footnotesize $\Big(C^1_{\frac{1}{2}},I_{(2,1)} \Big)
\longrightarrow \Big(C^1_{\frac{1}{2}},C^1_{\frac{1}{2}}.ii\Big)
$}

$$
{\footnotesize C=\left( \begin{tabular}{cccccc}
              $1$ & $a$& $e$& $b$ & $ 0$& $ 0$\\
              $0$ & $c^2$& $-bc^2$& $0$& $ 0$& $ 0$ \\
               $0$ & $-c^2$& $c^2d$& $0$& $0$& $0$ \\
               $1$ & $a$& $-a(d-b)+e$& $ d$& $ 0$& $ 0$ \\
                 $0$ & $0$& $ 0$& $0$& $c$& $bc$ \\
                 $0$ & $0$& $ 0$& $ 0$& $c$& $cd$ \\
                 \end{tabular} \right),\;\;\;\;b-d \neq 0;\;\;c\in\Re-\{0\};\;\;
                 a,b,d,e\in\Re.}
$$
%%%%%%%%%%%%%%%%%%%%%%%%%%%%%%%%%%%%%%%%%%%%%%%%%%%%%%%%%%%%%%%%%%%%%%%%%%%%%%%%%%%%%
%%%%%%%%%%%%%%%%%%%%%%%%%%%%%%%%%%%%%%%%%%%%%%%%%%%%%%%%%%%%%%%%%%%%%%%%%%%%%%%%%%%%%
\vspace{1.2cm}

Isomorphisms of the Drinfeld's superdoubles of the type $(2,4)$\\

\vspace{0.7cm}

{\bf $\cal{D}$$sd^{2}_{(2,4)}:$}\\

{\footnotesize $\Big(C^2_{0} ,
(A_{1,1}+2A)^0_{0,0,\epsilon_1}\Big)\longrightarrow \Big(C^2_{0} ,
(A_{1,1}+2A)^0_{\epsilon_2,\epsilon_2,\epsilon_2
}\Big),\;\;\epsilon_1, \epsilon_2=\pm 1 $}

$$
{\footnotesize C=\left( \begin{tabular}{cccccc}
              $1$ & $d$& $ 0$& $ 0$ & $ 0$& $ 0$\\
              $0$ & $\frac{\epsilon_1 a^2}{\epsilon_2}$& $ 0$& $ 0$& $ 0$& $ 0$ \\
              $0$ & $0$& $ \frac{\epsilon_1 a^2}{\epsilon_2 b}$& $ 0$& $0$& $ 0$ \\
              $0$ & $0$& $ 0$& $\frac{\epsilon_1 a}{\epsilon_2}$& $ 0$& $ 0$ \\
              $0$ & $0$& $ -\frac{\epsilon_1 a^2}{2b}$& $ -\epsilon_1 a$& $ b$& $ 0$ \\
              $0$ & $0$& $- \frac{\epsilon_1 a^2}{b}$& $c$& $0$& $a$ \\
                 \end{tabular} \right),\;\;\;\;\;\;
                 a, b\in\Re-\{0\};\;\;\;
                 c, d\in\Re,}
$$

\vspace{0.7cm}

{\footnotesize $\Big(C^2_{0} ,
(A_{1,1}+2A)^0_{0,0,\epsilon_1}\Big)\longrightarrow \Big(C^2_{0} ,
(A_{1,1}+2A)^1_{\epsilon_2,k,\epsilon_2 }\Big),\;\;\epsilon_1,
\epsilon_2=\pm 1;\;\;-1<k<1 $}

$$
{\footnotesize C=\left( \begin{tabular}{cccccc}
              $1$ & $d$& $ 0$& $ 0$ & $ 0$& $ 0$\\
              $0$ & $\frac{\epsilon_1 a^2}{\epsilon_2}$& $ 0$& $ 0$& $ 0$& $ 0$ \\
              $0$ & $0$& $ \frac{\epsilon_1 a^2}{\epsilon_2 b}$& $ 0$& $0$& $ 0$ \\
              $0$ & $0$& $ 0$& $\frac{\epsilon_1 a}{\epsilon_2}$& $ 0$& $ 0$ \\
              $0$ & $0$& $ -\frac{\epsilon_1 a^2}{2b}$& $ -\frac{\epsilon_1 k a}{\epsilon_2}$& $ b$& $ 0$ \\
              $0$ & $0$& $- \frac{\epsilon_1 k a^2}{\epsilon_2 b}$& $c$& $0$& $a$ \\
                 \end{tabular} \right),\;\;\;\;\;\;
                 a, b\in\Re-\{0\};\;\;\;
                 c, d\in\Re,}
$$

\vspace{0.7cm}

{\footnotesize $\Big(C^2_{0} ,
(A_{1,1}+2A)^0_{0,0,\epsilon_1}\Big)\longrightarrow \Big(C^2_{0} ,
(A_{1,1}+2A)^2_{0,1,\epsilon_2 }\Big),\;\;\epsilon_1,
\epsilon_2=\pm 1$}

$$
{\footnotesize C=\left( \begin{tabular}{cccccc}
              $1$ & $d$& $ 0$& $ 0$ & $ 0$& $ 0$\\
              $0$ & $\frac{\epsilon_1 a^2}{\epsilon_2}$& $ 0$& $ 0$& $ 0$& $ 0$ \\
              $0$ & $0$& $ -b$& $ 0$& $0$& $ 0$ \\
              $0$ & $0$& $ 0$& $\frac{\epsilon_1 a}{\epsilon_2}$& $ 0$& $ 0$ \\
              $0$ & $0$& $ 0$& $ -\frac{\epsilon_1  a}{\epsilon_2}$& $-\frac{\epsilon_1 a^2}{\epsilon_2b}$& $ 0$ \\
              $0$ & $0$& $b$& $c$& $0$& $a$ \\
                 \end{tabular} \right),\;\;\;\;\;\;
                 a, b\in\Re-\{0\};\;\;\;
                 c, d\in\Re,}
$$

\vspace{0.6cm}

{\footnotesize $\Big(C^2_{0} ,
(A_{1,1}+2A)^0_{0,0,\epsilon_1}\Big)\longrightarrow \Big(C^2_{0} ,
(A_{1,1}+2A)^2_{\epsilon_2,k,-\epsilon_2 }\Big),\;\;\epsilon_1,
\epsilon_2=\pm 1;\;\;k\in \Re$}

$$
{\footnotesize C=\left( \begin{tabular}{cccccc}
              $1$ & $d$& $ 0$& $ 0$ & $ 0$& $ 0$\\
              $0$ & $-\frac{\epsilon_1 a^2}{\epsilon_2}$& $ 0$& $ 0$& $ 0$& $ 0$ \\
              $0$ & $0$& $-\frac{\epsilon_1 a^2}{\epsilon_2b}$& $ 0$& $0$& $ 0$ \\
              $0$ & $0$& $ 0$& $-\frac{\epsilon_1 a}{\epsilon_2}$& $ 0$& $ 0$ \\
              $0$ & $0$&$-\frac{\epsilon_1 a^2}{2b}$&$\frac{\epsilon_1k a}{\epsilon_2}$&$b$& $ 0$ \\
              $0$ & $0$& $\frac{\epsilon_1k a^2}{\epsilon_2b}$& $c$& $0$& $a$ \\
                 \end{tabular} \right),\;\;\;\;\;\;
                 a, b\in\Re-\{0\};\;\;\;
                 c, d\in\Re,}
$$

\vspace{0.9cm}

{\bf $\cal{D}$$sd^{3\;\;p=0}_{(2,4)}:$\\

{\footnotesize $\Big(C^2_{0} , (A_{1,1}+2A)^0_{\epsilon,0,0
}\Big)\longrightarrow \Big(C^2_{0} , (A_{1,1}+2A)^2_{0,1,0
}\Big),\;\;\;\epsilon = \pm 1 $}

$$
{\footnotesize C=\left( \begin{tabular}{cccccc}
              $1$ & $e$& $ 0$& $ 0$ & $ 0$& $ 0$\\
              $0$ & $-ab$& $ 0$& $ 0$& $ 0$& $ 0$ \\
              $0$ & $0$& $ -a$& $ 0$& $0$& $ 0$ \\
              $0$ & $0$& $ 0$& $-c$& $0$& $ -d$ \\
              $0$ & $0$& $ \frac{\epsilon b}{2}$& $ c$& $b$& $d$ \\
              $0$ & $0$& $ a$& $h$& $0$& $ g$ \\
                 \end{tabular} \right),\;\;\;\;\;\;hd-gc \neq 0;\;\;a, b
                 \in\Re-\{0\};\;\;\;c, d,e, g, h \in\Re,}
$$

\vspace{0.7cm}

{\footnotesize $\Big(C^2_{0} , (A_{1,1}+2A)^0_{\epsilon_1 ,0,0
}\Big)\longrightarrow \Big(C^2_{0} ,
(A_{1,1}+2A)^2_{\epsilon_2,1,0 }\Big),\;\;\;\epsilon_1, \epsilon_2
= \pm 1 $}

$$
{\footnotesize C=\left( \begin{tabular}{cccccc}
              $1$ & $e$& $ 0$& $ 0$ & $ 0$& $ 0$\\
              $0$ & $-ab$& $ 0$& $ 0$& $ 0$& $ 0$ \\
              $0$ & $0$& $ -a$& $ 0$& $0$& $ 0$ \\
              $0$ & $0$& $ 0$& $-c$& $0$& $ -d$ \\
              $0$ & $0$& $ \frac{\epsilon_1 b+\epsilon_2 a}{2}$& $ c$& $b$& $d$ \\
              $0$ & $0$& $ a$& $h$& $0$& $ g$ \\
                 \end{tabular} \right),\;\;\;\;\;\;hd-gc \neq 0;\;\;a, b
                 \in\Re-\{0\};\;\;\;c, d,e, g, h \in\Re,}
$$

\vspace{0.7cm}

{\footnotesize $\Big(C^2_{0} , (A_{1,1}+2A)^0_{\epsilon ,0,0
}\Big)\longrightarrow \Big(C^2_{0} ,
I_{(1,2)}\Big),\;\;\;\epsilon = \pm 1 $}

$$
{\footnotesize C=\left( \begin{tabular}{cccccc}
              $1$ & $e$& $ 0$& $ 0$ & $ 0$& $ 0$\\
              $0$ & $ab$& $ 0$& $ 0$& $ 0$& $ 0$ \\
              $0$ & $0$& $ a$& $ 0$& $0$& $ 0$ \\
              $0$ & $0$& $ 0$& $c$& $0$& $ d$ \\
              $0$ & $0$& $ \frac{\epsilon b}{2}$& $ 0$& $b$& $0$ \\
              $0$ & $0$& $ 0$& $h$& $0$& $ g$ \\
                 \end{tabular} \right),\;\;\;\;\;\;hd-gc \neq 0;\;\;a, b
                 \in\Re-\{0\};\;\;\;c, d,e, g, h \in\Re,}
$$

\vspace{0.7cm}

{\bf $\cal{D}$$sd^{3\;\;p=1}_{(2,4)}:$}\\

{\footnotesize $\Big(C^2_1 ,
(A_{1,1}+2A)^0_{0,0,\epsilon_1}\Big)\longrightarrow \Big(C^2_{1} ,
(A_{1,1}+2A)^1_{\epsilon_2,0,\epsilon_2 }\Big),\;\;\;\epsilon_1,
\epsilon_2=\pm 1 $}

$$
{\footnotesize C=\left( \begin{tabular}{cccccc}
              $1$ & $c$& $ 0$& $ 0$ & $ 0$& $ 0$\\
              $0$ & $ab$& $ 0$& $ 0$& $ 0$& $ 0$ \\
              $0$ & $0$& $0$& $ \frac{ab}{d}$& $0$& $ 0$ \\
              $0$ & $0$& $a$& $-\frac{ae}{d}$& $ 0$& $ 0$ \\
              $0$ & $0$& $ 0$& $ \frac{\epsilon_1 d^2-\epsilon_2ab}{2d}$& $ e$& $d$ \\
              $0$ & $0$& $ -\epsilon_2\frac{ a}{2}$& $\epsilon_2\frac{ ae}{2d}$& $b$& $0$ \\
                 \end{tabular} \right),\;\;\;\;\;\;
                 a, b, d\in\Re-\{0\};\;\;\;
                 c, e\in\Re,}
$$

\vspace{0.60cm}

{\footnotesize $\Big(C^2_1 ,
(A_{1,1}+2A)^0_{0,0,\epsilon_1}\Big)\longrightarrow \Big(C^2_{1} ,
(A_{1,1}+2A)^2_{\epsilon_2,0,-\epsilon_2 }\Big),\;\;\;\epsilon_1,
\epsilon_2=\pm 1 $}

$$
{\footnotesize C=\left( \begin{tabular}{cccccc}
              $1$ & $e$& $ 0$& $ 0$ & $ 0$& $ 0$\\
              $0$ & $ab+cd$&$ 0$& $ 0$& $ 0$& $ 0$ \\
              $0$ & $0$& $-\frac{bc}{f}$& $ \frac{cd}{f}$& $0$& $ 0$ \\
              $0$ & $0$& $c$& $a$& $ 0$& $ 0$ \\
              $0$ & $0$& $ \frac{\epsilon_2 bc}{2f}$& $ \frac{\epsilon_1 f^2-\epsilon_2cd}{2f}$& $ -\frac{af}{c}$& $f$ \\
              $0$ & $0$& $\epsilon_2\frac{ c}{2}$& $ \frac{\epsilon_1 b+\epsilon_2a}{2}$& $d$& $b$ \\
                 \end{tabular} \right),\;\;\;\;\;\;ab+cd\neq
                 0;\;\; \;f, c\in\Re-\{0\};\;\;\;
                 a, b, d, e\in\Re,}
$$

\vspace{0.60cm}

{\footnotesize $\Big(C^2_1 ,
(A_{1,1}+2A)^0_{0,0,\epsilon_1}\Big)\longrightarrow
\Big(C^2_{p=-1} , (A_{1,1}+2A)^0_{\epsilon_2,0,0
}\Big),\;\;\;\epsilon_1, \epsilon_2=\pm 1 $}

$$
{\footnotesize C=\left( \begin{tabular}{cccccc}
              $1$ & $c$& $ 0$& $ 0$ & $ 0$& $ 0$\\
              $0$ & $-ab$&$ 0$& $ 0$& $ 0$& $ 0$ \\
              $0$ & $0$& $\frac{ad}{e}$& $ -\frac{ab}{e}$& $0$& $ 0$ \\
              $0$ & $0$& $0$& $\epsilon_1 \frac{d}{2}$& $b$& $d$ \\
              $0$ & $0$& $ -\epsilon_2 \frac{ad}{2e}$& $ \frac{\epsilon_1 e^2+\epsilon_2ab}{2e}$& $0$& $e$ \\
              $0$ & $0$& $f$& $0$& $0$& $0$ \\
                 \end{tabular} \right),\;\;\;\;\;\; \;a, b, e, f\in\Re-\{0\};\;\;\;
                 c, d\in\Re,}
$$

\vspace{0.60cm}

{\footnotesize $\Big(C^2_1 ,
(A_{1,1}+2A)^0_{0,0,\epsilon_1}\Big)\longrightarrow
\Big(C^2_{p=-1} , (A_{1,1}+2A)^0_{0,0,\epsilon_2
}\Big),\;\;\;\epsilon_1, \epsilon_2=\pm 1 $}

$$
{\footnotesize C=\left( \begin{tabular}{cccccc}
              $-1$ & $c$& $ 0$& $ 0$ & $ 0$& $ 0$\\
              $0$ & $-ab$&$ 0$& $ 0$& $ 0$& $ 0$ \\
              $0$ & $0$& $0$& $ -\epsilon_1\frac{ab}{2d}$& $\frac{eb}{d}$& $-\frac{ab}{d}$ \\
              $0$ & $0$& $a$& $e$& $0$& $0$ \\
              $0$ & $0$& $0$& $d$& $0$& $0$ \\
              $0$ & $0$& $\epsilon_2 \frac{a}{2}$& $\epsilon_2 \frac{e}{2}$& $b$& $0$ \\
                 \end{tabular} \right),\;\;\;\;\;\; \;a, b, d\in\Re-\{0\};\;\;\;
                 c, e\in\Re,}
$$

\vspace{0.60cm}

{\footnotesize $\Big(C^2_1 ,
(A_{1,1}+2A)^0_{0,0,\epsilon_1}\Big)\longrightarrow
\Big(C^2_{p=-1} , (A_{1,1}+2A)^1_{\epsilon_2,k=0,\epsilon_2
}\Big),\;\;\;\epsilon_1, \epsilon_2=\pm 1 $}

$$
{\footnotesize C=\left( \begin{tabular}{cccccc}
              $-1$ & $c$& $ 0$& $ 0$ & $ 0$& $ 0$\\
              $0$ & $-ab$&$ 0$& $ 0$& $ 0$& $ 0$ \\
              $0$ & $0$& $0$& $0$& $-\frac{ab}{d}$& $0$ \\
              $0$ & $0$& $0$& $a$& $0$& $0$ \\
              $0$ & $0$& $d$& $ -\frac{de}{b}$& $\epsilon_2 \frac{ab}{2d}$& $0$ \\
              $0$ & $0$& $0$& $ \frac{\epsilon_1 b+\epsilon_2 a}{2}$& $e$& $b$ \\
                 \end{tabular} \right),\;\;\;\;\;\; \;a, b, d\in\Re-\{0\};\;\;\;
                 c, e\in\Re,}
$$

\vspace{0.60cm}

{\footnotesize $\Big(C^2_1 ,
(A_{1,1}+2A)^0_{0,0,\epsilon_1}\Big)\longrightarrow
\Big(C^2_{p=-1} , (A_{1,1}+2A)^2_{\epsilon_2,s=0,-\epsilon_2
}\Big),\;\;\;\epsilon_1, \epsilon_2=\pm 1 $}

$$
{\footnotesize C=\left( \begin{tabular}{cccccc}
              $-1$ & $c$& $ 0$& $ 0$ & $ 0$& $ 0$\\
              $0$ & $a\epsilon_2(2b-\epsilon_1a)$&$ 0$& $ 0$& $ 0$& $ 0$ \\
              $0$ & $0$& $0$& $0$& $-2\epsilon_2 d$& $0$ \\
              $0$ & $0$& $0$& $-\epsilon_2(2b-\epsilon_1a)$& $0$& $0$ \\
              $0$ & $0$& $-\frac{a(2b-\epsilon_1a)}{2d}$& $\frac{e(2b-\epsilon_1a)}{2d}$& $d$& $0$ \\
              $0$ & $0$& $0$& $b$& $e$& $a$ \\
                 \end{tabular} \right),\;\;\;\;2b-\epsilon_1a\neq 0;\;\;a, d\in\Re-\{0\};\;\;\;
                 b, c, e\in\Re,}
$$

\vspace{0.60cm}

{\footnotesize $\Big(C^2_1 ,
(A_{1,1}+2A)^0_{0,0,\epsilon}\Big)\longrightarrow \Big(C^2_{1} ,
I_{(1,2)}\Big),\;\;\;\epsilon=\pm 1 $}

$$
{\footnotesize C=\left( \begin{tabular}{cccccc}
              $1$ & $c$& $ 0$& $ 0$ & $ 0$& $ 0$\\
              $0$ & $ab$&$ 0$& $ 0$& $ 0$& $ 0$ \\
              $0$ & $0$& $\frac{ab}{d}$& $-\frac{ae}{d}$& $0$& $0$ \\
              $0$ & $0$& $0$& $a$& $0$& $0$ \\
              $0$ & $0$& $0$& $0$& $d$& $0$ \\
              $0$ & $0$& $0$& $\epsilon \frac{b}{2}$& $e$& $b$ \\
                 \end{tabular} \right),\;\;\;\;\;\;\;a, b, d\in\Re-\{0\};\;\;\;
                  c, e\in\Re,}
$$

\vspace{0.60cm}

{\footnotesize $\Big(C^2_1 ,
(A_{1,1}+2A)^0_{0,0,\epsilon}\Big)\longrightarrow \Big(C^2_{p=-1}
, I_{(1,2)}\Big),\;\;\;\epsilon=\pm 1 $}

$$
{\footnotesize C=\left( \begin{tabular}{cccccc}
              $1$ & $c$& $ 0$& $ 0$ & $ 0$& $ 0$\\
              $0$ & $-ab$&$ 0$& $ 0$& $ 0$& $ 0$ \\
              $0$ & $0$& $\frac{ae}{d}$& $-\frac{ab}{d}$& $0$& $0$ \\
              $0$ & $0$& $0$& $\epsilon \frac{e}{2}$& $a$& $e$ \\
              $0$ & $0$& $0$& $\epsilon \frac{d}{2}$& $0$& $d$ \\
              $0$ & $0$& $b$& $0$& $0$& $0$ \\
                 \end{tabular} \right),\;\;\;\;\;\;\;a, b, d\in\Re-\{0\};\;\;\;
                  c, e\in\Re,}
$$

\vspace{0.70cm}

{\bf $\cal{D}$$sd^{3\;\;p}_{(2,4)}:\;\;\;\;\;p\in (-1,1)-\{0\}$}\\

{\footnotesize $\Big(C^2_p ,
(A_{1,1}+2A)^0_{\epsilon_1,0,0}\Big)\longrightarrow \Big(C^2_{-p}
,
(A_{1,1}+2A)^0_{\epsilon_2,0,0}\Big),\;\;\;\epsilon_1,\;\epsilon_2=\pm
1 $}

$$
{\footnotesize C=\left( \begin{tabular}{cccccc}
              $1$ & $d$& $ 0$& $ 0$ & $ 0$& $ 0$\\
              $0$ & $-ab$&$ 0$& $ 0$& $ 0$& $ 0$ \\
              $0$ & $0$& $\frac{-ab}{c}$& $0$& $0$& $0$ \\
              $0$ & $0$& $0$& $0$& $0$& $a$ \\
              $0$ & $0$& $ \frac{\epsilon_1 c^2+\epsilon_2ab}{2c}$& $0$& $c$& $0$ \\
              $0$ & $0$& $0$& $b$& $0$& $0$ \\
                 \end{tabular} \right),\;\;\;\;\;\;\;a, b, c\in\Re-\{0\};\;\;\;
                  d\in\Re,}
$$

\vspace{0.60cm}

{\footnotesize $\Big(C^2_p ,
(A_{1,1}+2A)^0_{\epsilon_1,0,0}\Big)\longrightarrow \Big(C^2_{p} ,
(A_{1,1}+2A)^0_{0,0,\epsilon_2}\Big),\;\;\;\epsilon_1,\;\epsilon_2=\pm
1 $}

$$
{\footnotesize C=\left( \begin{tabular}{cccccc}
              $1$ & $d$& $ 0$& $ 0$ & $ 0$& $ 0$\\
              $0$ & $ab$&$ 0$& $ 0$& $ 0$& $ 0$ \\
              $0$ & $0$& $\frac{ab}{c}$& $0$& $0$& $0$ \\
              $0$ & $0$& $0$& $a$& $0$& $0$ \\
              $0$ & $0$& $\frac{\epsilon_1 c}{2}$& $0$& $c$& $0$ \\
              $0$ & $0$& $ 0$& $-\frac{\epsilon_2 a}{2p}$& $0$& $b$ \\
                 \end{tabular} \right),\;\;\;\;\;\;\;a, b, c\in\Re-\{0\};\;\;\;
                  d\in\Re,}
$$

\vspace{0.60cm}

{\footnotesize $\Big(C^2_p ,
(A_{1,1}+2A)^0_{\epsilon_1,0,0}\Big)\longrightarrow \Big(C^2_{p} ,
(A_{1,1}+2A)^0_{\epsilon_2,\epsilon_2,\epsilon_2}\Big),\;\;\;\epsilon_1,\;\epsilon_2=\pm
1 $}

$$
{\footnotesize C=\left( \begin{tabular}{cccccc}
              $1$ & $d$& $ 0$& $ 0$ & $ 0$& $ 0$\\
              $0$ & $-\frac{(p+1)ab}{\epsilon_2}$&$ 0$& $ 0$& $ 0$& $ 0$ \\
              $0$ & $0$& $-\frac{(p+1)a}{\epsilon_2}$& $0$& $0$& $0$ \\
              $0$ & $0$& $0$& $-\frac{(p+1)ab}{\epsilon_2c}$& $0$& $0$ \\
              $0$ & $0$& $\frac{\epsilon_1 b+(1+p)a}{2}$& $\frac{ab}{c}$& $b$& $0$ \\
              $0$ & $0$& $ a$& $\frac{(p+1)ab}{2pc}$& $0$& $c$ \\
                 \end{tabular} \right),\;\;\;\;\;\;\;a, b, c\in\Re-\{0\};\;\;\;
                  d\in\Re,}
$$

\vspace{0.60cm}

{\footnotesize $\Big(C^2_p ,
(A_{1,1}+2A)^0_{\epsilon_1,0,0}\Big)\longrightarrow \Big(C^2_{p} ,
(A_{1,1}+2A)^1_{\epsilon_2,k,\epsilon_2}\Big),\;\;\;\;-1<k<1;\;\;\epsilon_1,\;\epsilon_2=\pm
1 $}

$$
{\footnotesize C=\left( \begin{tabular}{cccccc}
              $1$ & $d$& $ 0$& $ 0$ & $ 0$& $ 0$\\
              $0$ & $ab$&$ 0$& $ 0$& $ 0$& $ 0$ \\
              $0$ & $0$& $a$& $0$& $0$& $0$ \\
              $0$ & $0$& $0$& $\frac{ab}{c}$& $0$& $0$ \\
              $0$ & $0$& $\frac{\epsilon_1 b- \epsilon_2 a}{2}$& $\frac{-kab}{c(p+1)}$& $b$& $0$ \\
              $0$ & $0$& $ \frac{-ka}{p+1}$& $-\frac{\epsilon_2ab}{2pc}$& $0$& $c$ \\
                 \end{tabular} \right),\;\;\;\;\;\;\;a, b, c\in\Re-\{0\};\;\;\;
                  d\in\Re,}
$$

\vspace{0.60cm}

{\footnotesize $\Big(C^2_p ,
(A_{1,1}+2A)^0_{\epsilon,0,0}\Big)\longrightarrow \Big(C^2_{p} ,
(A_{1,1}+2A)^2_{0,1,0}\Big),\;\;\;\;\epsilon=\pm 1 $}

$$
{\footnotesize C=\left( \begin{tabular}{cccccc}
              $1$ & $d$& $ 0$& $ 0$ & $ 0$& $ 0$\\
              $0$ & $-ab(1+p)$&$ 0$& $ 0$& $ 0$& $ 0$ \\
              $0$ & $0$& $-a(1+p)$& $0$& $0$& $0$ \\
              $0$ & $0$& $0$& $-\frac{ab(1+p)}{c}$& $0$& $0$ \\
              $0$ & $0$& $\frac{\epsilon b}{2}$& $\frac{ab}{c}$& $b$& $0$ \\
              $0$ & $0$& $a$& $0$& $0$& $c$ \\
                 \end{tabular} \right),\;\;\;\;\;\;\;a, b, c\in\Re-\{0\};\;\;\;
                  d\in\Re,}
$$

\vspace{0.60cm}

{\footnotesize $\Big(C^2_p ,
(A_{1,1}+2A)^0_{\epsilon_1,0,0}\Big)\longrightarrow \Big(C^2_{p} ,
(A_{1,1}+2A)^2_{\epsilon_2,1,0}\Big),\;\;\;\;\epsilon_1,\;\epsilon_2=\pm
1 $}

$$
{\footnotesize C=\left( \begin{tabular}{cccccc}
              $1$ & $d$& $ 0$& $ 0$ & $ 0$& $ 0$\\
              $0$ & $-ab(1+p)$&$ 0$& $ 0$& $ 0$& $ 0$ \\
              $0$ & $0$& $-a(1+p)$& $0$& $0$& $0$ \\
              $0$ & $0$& $0$& $-\frac{ab(1+p)}{c}$& $0$& $0$ \\
              $0$ & $0$& $\frac{\epsilon_1 b+a\epsilon_2(1+p)}{2}$& $\frac{ab}{c}$& $b$& $0$ \\
              $0$ & $0$& $a$& $0$& $0$& $c$ \\
                 \end{tabular} \right),\;\;\;\;\;\;\;a, b, c\in\Re-\{0\};\;\;\;
                  d\in\Re,}
$$

\vspace{0.60cm}

{\footnotesize $\Big(C^2_p ,
(A_{1,1}+2A)^0_{\epsilon_1,0,0}\Big)\longrightarrow \Big(C^2_{p} ,
(A_{1,1}+2A)^2_{0,1,\epsilon_2}\Big),\;\;\;\;\epsilon_1,\;\epsilon_2=\pm
1 $}

$$
{\footnotesize C=\left( \begin{tabular}{cccccc}
              $1$ & $d$& $ 0$& $ 0$ & $ 0$& $ 0$\\
              $0$ & $-ab(1+p)$&$ 0$& $ 0$& $ 0$& $ 0$ \\
              $0$ & $0$& $-a(1+p)$& $0$& $0$& $0$ \\
              $0$ & $0$& $0$& $-\frac{ab(1+p)}{c}$& $0$& $0$ \\
              $0$ & $0$& $\frac{\epsilon_1 b}{2}$& $\frac{ab}{c}$& $b$& $0$ \\
              $0$ & $0$& $a$& $\frac{\epsilon_2 ab(1+p)}{2pc}$& $0$& $c$ \\
                 \end{tabular} \right),\;\;\;\;\;\;\;a, b, c\in\Re-\{0\};\;\;\;
                  d\in\Re,}
$$

\vspace{0.60cm}

{\footnotesize $\Big(C^2_p ,
(A_{1,1}+2A)^0_{\epsilon_1,0,0}\Big)\longrightarrow \Big(C^2_{p} ,
(A_{1,1}+2A)^2_{\epsilon_2,s,-\epsilon_2}\Big),\;\;\;\;\;\;s\in
\Re;\;\;\; \epsilon_1,\;\epsilon_2=\pm 1 $}

$$
{\footnotesize C=\left( \begin{tabular}{cccccc}
              $1$ & $d$& $ 0$& $ 0$ & $ 0$& $ 0$\\
              $0$ & $a\epsilon_2(\epsilon_1 a-2b)$&$ 0$& $ 0$& $ 0$& $ 0$ \\
              $0$ & $0$& $\epsilon_2(\epsilon_1 a-2b)$& $0$& $0$& $0$ \\
              $0$ & $0$& $0$& $c$& $0$& $0$ \\
              $0$ & $0$& $b$& $-\frac{sc}{p+1}$& $a$& $0$ \\
              $0$ & $0$& $-\frac{s\epsilon_2 (\epsilon_1 a-2b)}{p+1}$& $\frac{\epsilon_2 c}{2p}$& $0$& $\frac{a\epsilon_2 (\epsilon_1 a-2b)}{c}$ \\
                 \end{tabular} \right),\;\;\;\;\;\;\;\epsilon_1 a-2b\neq 0;\;\; a, c\in\Re-\{0\};\;\;\;
                  b, d\in\Re,}
$$

\vspace{0.60cm}

{\footnotesize $\Big(C^2_p ,
(A_{1,1}+2A)^0_{\epsilon,0,0}\Big)\longrightarrow \Big(C^2_{p} ,
I_{(1,2)}\Big),\;\;\;\;\;\epsilon=\pm 1 $}

$$
{\footnotesize C=\left( \begin{tabular}{cccccc}
              $1$ & $d$& $ 0$& $ 0$ & $ 0$& $ 0$\\
              $0$ & $ab$&$ 0$& $ 0$& $ 0$& $ 0$ \\
              $0$ & $0$& $\frac{ab}{c}$& $0$& $0$& $0$ \\
              $0$ & $0$& $0$& $a$& $0$& $0$ \\
              $0$ & $0$& $\frac{\epsilon c}{2}$& $0$& $c$& $0$ \\
              $0$ & $0$& $0$& $0$& $0$& $b$ \\
                 \end{tabular} \right),\;\;\;\;\;\;\;\; a, b, c\in\Re-\{0\};\;\;\;
                  d\in\Re,}
$$

\vspace{0.60cm}

{\bf $\cal{D}$$sd^{4}_{(2,4)}:$}\\

{\footnotesize $\Big(C^3 ,
(A_{1,1}+2A)^0_{1,0,0}\Big)\longrightarrow \Big(C^3 ,
(A_{1,1}+2A)^2_{0,\epsilon,0 }\Big),\;\;\;\epsilon=\pm 1 $}

$$
{\footnotesize C=\left( \begin{tabular}{cccccc}
              $a$ & $b$& $ 0$& $ 0$ & $ 0$& $ 0$\\
              $0$ & $2cd$& $ 0$& $ 0$& $ 0$& $ 0$ \\
              $0$ & $0$& $ac$& $0$& $0$& $ 0$ \\
              $0$ & $0$& $e$& $c$& $ 0$& $ f$ \\
              $0$ & $0$& $ g$& $d$& $2d$& $h$ \\
              $0$ & $0$& $ad-e \epsilon$& $-\epsilon c$& $0$& $2ad-\epsilon f$ \\
                 \end{tabular} \right),\;\;\;\;\;\;
                 a, c, d\in\Re-\{0\};\;\;\;
                 b, e, f, g, h\in\Re,}
$$

\vspace{0.60cm}

{\footnotesize $\Big(C^3 ,
(A_{1,1}+2A)^0_{1,0,0}\Big)\longrightarrow \Big(C^3 ,
I_{(1,2)}\Big)$}

$$
{\footnotesize C=\left( \begin{tabular}{cccccc}
              $a$ & $b$& $ 0$& $ 0$ & $ 0$& $ 0$\\
              $0$ & $cd$& $ 0$& $ 0$& $ 0$& $ 0$ \\
              $0$ & $0$& $ac$& $0$& $0$& $ 0$ \\
              $0$ & $0$& $e$& $c$& $ 0$& $ f$ \\
              $0$ & $0$& $ g$& $ \frac{d}{2}$& $d$& $h$ \\
              $0$ & $0$& $\frac{ ad}{2}$& $0$& $0$& $ad$ \\
                 \end{tabular} \right),\;\;\;\;\;\;
                 a, c, d\in\Re-\{0\};\;\;\;
                 b, e, f, g, h\in\Re,}
$$

\vspace{0.60cm}

{\footnotesize $\Big(C^3 ,
(A_{1,1}+2A)^0_{1,0,0}\Big)\longrightarrow \Big((A_{1,1}+2A)^2 ,
I_{(1,2)}\Big)$}

$$
{\footnotesize C=\left( \begin{tabular}{cccccc}
              $0$ & $c(c-2b)$& $ 0$& $ 0$ & $ 0$& $ 0$\\
              $a$ & $d$& $ 0$& $ 0$& $ 0$& $ 0$ \\
              $0$ & $0$& $e$& $b$& $c$& $ g$ \\
              $0$ & $0$& $f$& $c-b$& $ c$& $ h$ \\
              $0$ & $0$& $ a(b-c)$& $ 0$& $0$& $-ac$ \\
              $0$ & $0$& $ab$& $0$& $0$& $ac$ \\
                 \end{tabular} \right),\;\;\;\;\;c-2b \neq 0;\;\;
                 a, c\in\Re-\{0\};\;\;\;
                 b, d, e, f, g, h\in\Re,}
$$

\vspace{0.70cm}

{\bf $\cal{D}$$sd^{5}_{(2,4)}:$}\\

{\footnotesize $\Big(C^3 ,
(A_{1,1}+2A)^0_{0,0,1}\Big)\longrightarrow \Big(C^3 ,
(A_{1,1}+2A)^1_{\epsilon,0,\epsilon }\Big),\;\;\;\epsilon=\pm 1 $}

$$
{\footnotesize C=\left( \begin{tabular}{cccccc}
              $a$ & $b$& $ 0$& $ 0$ & $ 0$& $ 0$\\
              $0$ & $\epsilon\frac{c^2}{a^2}$& $ 0$& $ 0$& $ 0$& $ 0$ \\
              $0$ & $0$& $ac$& $0$& $0$& $ 0$ \\
              $0$ & $0$& $d$& $c$& $ 0$& $ 0$ \\
              $0$ & $0$& $e$& $\epsilon \frac{d^2-a^2c^2}{2a^2c}$& $\epsilon\frac{c}{a^2}$& $-\epsilon\frac{d}{a^2}$ \\
              $0$ & $0$& $-\epsilon\frac{a^2c^2+d^2}{2ac}$& $-\epsilon \frac{d}{a}$& $0$& $\epsilon \frac{c}{a}$ \\
                 \end{tabular} \right),\;\;\;\;\;\;
                 a, c\in\Re-\{0\};\;\;\;
                 b, d, e\in\Re,}
$$

\vspace{0.60cm}

{\footnotesize $\Big(C^3 ,
(A_{1,1}+2A)^0_{0,0,1}\Big)\longrightarrow \Big(C^3 ,
(A_{1,1}+2A)^2_{\epsilon,0,-\epsilon }\Big),\;\;\;\epsilon=\pm 1
$}

$$
{\footnotesize C=\left( \begin{tabular}{cccccc}
              $a$ & $b$& $ 0$& $ 0$ & $ 0$& $ 0$\\
              $0$ & $-\epsilon a^2 c^2$& $ 0$& $ 0$& $ 0$& $ 0$ \\
              $0$ & $0$& $-\epsilon a^3 c$& $0$& $0$& $ 0$ \\
              $0$ & $0$& $\epsilon a^2 d$& $-\epsilon a^2 c$& $ 0$& $ 0$ \\
              $0$ & $0$& $e$& $\frac{d^2+a^2c^2}{2c}$& $c$& $d$ \\
              $0$ & $0$& $\frac{a(a^2c^2-d^2)}{2c}$& $ad$& $0$& $ac$ \\
                 \end{tabular} \right),\;\;\;\;\;\;
                 a, c\in\Re-\{0\};\;\;\;
                 b, d, e\in\Re,}
$$

\vspace{0.60cm}

{\bf $\cal{D}$$sd^{6}_{(2,4)}:$}\\

{\footnotesize $\Big(C^4 ,
(A_{1,1}+2A)^0_{\epsilon_1,0,0}\Big)\longrightarrow \Big(C^4 ,
(A_{1,1}+2A)^0_{0,0,\epsilon_2 }\Big),\;\;\;\epsilon_1,
\epsilon_2 =\pm 1 $}

$$
{\footnotesize C=\left( \begin{tabular}{cccccc}
              $1$ & $b$& $ 0$& $ 0$ & $ 0$& $ 0$\\
              $0$ & $ac$& $ 0$& $ 0$& $ 0$& $ 0$ \\
              $0$ & $0$& $a$& $0$& $0$& $ 0$ \\
              $0$ & $0$& $-\frac{ad}{c}$& $a$& $ 0$& $ 0$ \\
              $0$ & $0$& $\frac{2\epsilon_1 c^2-a\epsilon_2(c+d)}{4c}$& $\epsilon_2\frac{ a}{4}$& $c$& $d$ \\
              $0$ & $0$& $\frac{a\epsilon_2(c+2d)}{4c}$& $-\epsilon_2 \frac{a}{2}$& $0$& $c$ \\
                 \end{tabular} \right),\;\;\;\;\;\;
                 a, c\in\Re-\{0\};\;\;\;
                 b, d\in\Re,}
$$

\vspace{0.60cm}

{\footnotesize $\Big(C^4 ,
(A_{1,1}+2A)^0_{\epsilon_1,0,0}\Big)\longrightarrow \Big(C^4 ,
(A_{1,1}+2A)^1_{m,0,\epsilon_2 }\Big),\;\;\;\epsilon_1,
\epsilon_2 =\pm 1 $}

$$
{\footnotesize C=\left( \begin{tabular}{cccccc}
              $1$ & $c$& $ 0$& $ 0$ & $ 0$& $ 0$\\
              $0$ & $ab$& $ 0$& $ 0$& $ 0$& $ 0$ \\
              $0$ & $0$& $a$& $0$& $0$& $ 0$ \\
              $0$ & $0$& $-\frac{ad}{b}$& $a$& $ 0$& $ 0$ \\
              $0$ & $0$& $-\frac{a\epsilon_2(b+d)+2b(ma-\epsilon_1b)}{4b}$& $\epsilon_2\frac{a}{4}$& $b$& $d$ \\
              $0$ & $0$& $\frac{a\epsilon_2(b+2d)}{4b}$& $-\epsilon_2 \frac{a}{2}$& $0$& $b$ \\
                 \end{tabular} \right),\;\;\;\;\;\;
                 a, b\in\Re-\{0\};\;\;\;
                 c, d\in\Re,}
$$

\vspace{0.60cm}

{\footnotesize $\Big(C^4 ,
(A_{1,1}+2A)^0_{\epsilon_1,0,0}\Big)\longrightarrow \Big(C^4 ,
(A_{1,1}+2A)^2_{n,0,\epsilon_2 }\Big),\;\;\;\epsilon_1,
\epsilon_2 =\pm 1 $}

$$
{\footnotesize C=\left( \begin{tabular}{cccccc}
              $1$ & $c$& $ 0$& $ 0$ & $ 0$& $ 0$\\
              $0$ & $ab$& $ 0$& $ 0$& $ 0$& $ 0$ \\
              $0$ & $0$& $a$& $0$& $0$& $ 0$ \\
              $0$ & $0$& $-\frac{ad}{b}$& $a$& $ 0$& $ 0$ \\
              $0$ & $0$& $-\frac{a\epsilon_2(b+d)+2b(na-\epsilon_1b)}{4b}$& $\epsilon_2\frac{a}{4}$& $b$& $d$ \\
              $0$ & $0$& $\frac{a\epsilon_2(b+2d)}{4b}$& $-\epsilon_2 \frac{a}{2}$& $0$& $b$ \\
                 \end{tabular} \right),\;\;\;\;\;\;
                 a, b\in\Re-\{0\};\;\;\;
                 c, d\in\Re,}
$$

\vspace{0.60cm}

{\footnotesize $\Big(C^4 ,
(A_{1,1}+2A)^0_{\epsilon_1,0,0}\Big)\longrightarrow \Big(C^4 ,
(A_{1,1}+2A)^2_{0,\epsilon_2,0 }\Big),\;\;\;\epsilon_1, \epsilon_2
=\pm 1 $}

$$
{\footnotesize C=\left( \begin{tabular}{cccccc}
              $1$ & $c$& $ 0$& $ 0$ & $ 0$& $ 0$\\
              $0$ & $ab$& $ 0$& $ 0$& $ 0$& $ 0$ \\
              $0$ & $0$& $a$& $0$& $0$& $ 0$ \\
              $0$ & $0$& $-\frac{ad}{b}$& $a$& $ 0$& $ 0$ \\
              $0$ & $0$& $\frac{a\epsilon_2(b+d)+\epsilon_1 b^2}{2b}$& $-\epsilon_2\frac{a}{2}$& $b$& $d$ \\
              $0$ & $0$& $-\epsilon_2\frac{ a}{2}$& $0$& $0$& $b$ \\
                 \end{tabular} \right),\;\;\;\;\;\;
                 a, b\in\Re-\{0\};\;\;\;
                 c, d\in\Re,}
$$

\vspace{0.60cm}

{\footnotesize $\Big(C^4 ,
(A_{1,1}+2A)^0_{\epsilon,0,0}\Big)\longrightarrow \Big(C^4 ,
I_{(1,2) }\Big),\;\;\;\;\epsilon=\pm 1 $}

$$
{\footnotesize C=\left( \begin{tabular}{cccccc}
              $1$ & $c$& $ 0$& $ 0$ & $ 0$& $ 0$\\
              $0$ & $ab$& $ 0$& $ 0$& $ 0$& $ 0$ \\
              $0$ & $0$& $a$& $0$& $0$& $ 0$ \\
              $0$ & $0$& $-\frac{ad}{b}$& $a$& $ 0$& $ 0$ \\
              $0$ & $0$& $\epsilon\frac{b}{2}$& $0$& $b$& $d$ \\
              $0$ & $0$& $0$& $0$& $0$& $b$ \\
                 \end{tabular} \right),\;\;\;\;\;\;
                 a, b\in\Re-\{0\};\;\;\;
                 c, d\in\Re,}
$$

\vspace{0.60cm}

{\footnotesize $\Big(C^4 ,
(A_{1,1}+2A)^0_{\epsilon_1,0,0}\Big)\longrightarrow
\Big(C^2_{p=-1} , (A_{1,1}+2A)^1_{\epsilon_2,k,\epsilon_2
}\Big),\;\;\;\;k\in (-1,1)-\{0\};\;\;\; \epsilon_1, \epsilon_2
=\pm 1 $}

$$
{\footnotesize C=\left( \begin{tabular}{cccccc}
              $1$ & $c$& $ 0$& $ 0$ & $ 0$& $ 0$\\
              $0$ & $kab$& $ 0$& $ 0$& $ 0$& $ 0$ \\
              $0$ & $0$& $a$& $0$& $0$& $0$ \\
              $0$ & $0$& $0$& $0$& $ 0$& $b$ \\
              $0$ & $0$& $\frac{\epsilon_1 kb-\epsilon_2 a}{2}$& $0$& $kb$& $\frac{bd}{a}$ \\
              $0$ & $0$& $d$& $-ka$& $0$& $\epsilon_2\frac{b}{2}$ \\
                 \end{tabular} \right),\;\;\;\;\;\;
                 a, b\in\Re-\{0\};\;\;\;
                 c, d\in\Re,}
$$

\vspace{0.60cm}

{\footnotesize $\Big(C^4 ,
(A_{1,1}+2A)^0_{\epsilon_1,0,0}\Big)\longrightarrow
\Big(C^2_{p=-1} , (A_{1,1}+2A)^0_{\epsilon_2,\epsilon_2,\epsilon_2
}\Big),\;\;\;\;\; \epsilon_1, \epsilon_2 =\pm 1 $}

$$
{\footnotesize C=\left( \begin{tabular}{cccccc}
              $1$ & $c$& $ 0$& $ 0$ & $ 0$& $ 0$\\
              $0$ & $\epsilon_2ab$& $ 0$& $ 0$& $ 0$& $ 0$ \\
              $0$ & $0$& $a$& $0$& $0$& $0$ \\
              $0$ & $0$& $0$& $0$& $ 0$& $b$ \\
              $0$ & $0$& $\frac{\epsilon_2}{2} (\epsilon_1 b-a)$& $0$& $\epsilon_2b$& $\frac{bd}{a}$ \\
              $0$ & $0$& $d$& $-\epsilon_2a$& $0$& $\epsilon_2\frac{b}{2}$ \\
                 \end{tabular} \right),\;\;\;\;\;\;
                 a, b\in\Re-\{0\};\;\;\;
                 c, d\in\Re,}
$$

\vspace{0.60cm}

{\footnotesize $\Big(C^4 ,
(A_{1,1}+2A)^0_{\epsilon,0,0}\Big)\longrightarrow \Big(C^2_{p=-1}
, (A_{1,1}+2A)^2_{0,1,0}\Big),\;\;\;\;\;\; \epsilon =\pm 1 $}

$$
{\footnotesize C=\left( \begin{tabular}{cccccc}
              $1$ & $c$& $ 0$& $ 0$ & $ 0$& $ 0$\\
              $0$ & $-ab$& $ 0$& $ 0$& $ 0$& $ 0$ \\
              $0$ & $0$& $-b$& $0$& $0$& $0$ \\
              $0$ & $0$& $0$& $0$& $ 0$& $a$ \\
              $0$ & $0$& $\frac{\epsilon a}{2}$& $0$& $a$& $-\frac{ad}{b}$ \\
              $0$ & $0$& $d$& $b$& $0$& $0$ \\
                 \end{tabular} \right),\;\;\;\;\;\;
                 a, b\in\Re-\{0\};\;\;\;
                 c, d\in\Re,}
$$

\vspace{0.60cm}

{\footnotesize $\Big(C^4 ,
(A_{1,1}+2A)^0_{\epsilon_1,0,0}\Big)\longrightarrow
\Big(C^2_{p=-1} , (A_{1,1}+2A)^2_{\epsilon_2,1,0}\Big),\;\;\;\;\;
\epsilon_1, \epsilon_2 =\pm 1 $}

$$
{\footnotesize C=\left( \begin{tabular}{cccccc}
              $-1$ & $c$& $ 0$& $ 0$ & $ 0$& $ 0$\\
              $0$ & $-ab$& $ 0$& $ 0$& $ 0$& $ 0$ \\
              $0$ & $0$& $0$& $0$& $0$& $-b$ \\
              $0$ & $0$& $a$& $0$& $ 0$& $0$ \\
              $0$ & $0$& $-\frac{ad}{b}$& $a$& $0$& $\epsilon_2\frac{b}{2}$ \\
              $0$ & $0$& $\epsilon_1 \frac{b}{2}$& $0$& $b$& $d$ \\
                 \end{tabular} \right),\;\;\;\;\;\;
                 a, b\in\Re-\{0\};\;\;\;
                 c, d\in\Re,}
$$

\vspace{0.60cm}

{\footnotesize $\Big(C^4 ,
(A_{1,1}+2A)^0_{\epsilon_1,0,0}\Big)\longrightarrow
\Big(C^2_{p=-1} , (A_{1,1}+2A)^2_{0,1,\epsilon_2}\Big),\;\;\;\;\;
\epsilon_1, \epsilon_2 =\pm 1 $}

$$
{\footnotesize C=\left( \begin{tabular}{cccccc}
              $1$ & $c$& $ 0$& $ 0$ & $ 0$& $ 0$\\
              $0$ & $-ab$& $ 0$& $ 0$& $ 0$& $ 0$ \\
              $0$ & $0$& $-b$& $0$& $0$& $0$ \\
              $0$ & $0$& $0$& $0$& $ 0$& $a$ \\
              $0$ & $0$& $\epsilon_1 \frac{a}{2}$& $0$& $a$& $-\frac{ad}{b}$ \\
              $0$ & $0$& $d$& $b$& $0$& $\epsilon_2\frac{a}{2}$ \\
                 \end{tabular} \right),\;\;\;\;\;\;
                 a, b\in\Re-\{0\};\;\;\;
                 c, d\in\Re,}
$$

\vspace{0.60cm}

{\footnotesize $\Big(C^4 ,
(A_{1,1}+2A)^0_{\epsilon_1,0,0}\Big)\longrightarrow
\Big(C^2_{p=-1} ,
(A_{1,1}+2A)^2_{\epsilon_2,s,-\epsilon_2}\Big),\;\;\;\;\;\;s\in
\Re-\{0\};\;\;\; \;\epsilon_1, \epsilon_2 =\pm 1 $}

$$
{\footnotesize C=\left( \begin{tabular}{cccccc}
              $-1$ & $c$& $ 0$& $ 0$ & $ 0$& $ 0$\\
              $0$ & $-2sab \epsilon_2$& $ 0$& $ 0$& $ 0$& $ 0$ \\
              $0$ & $0$& $0$& $0$& $0$& $-2b \epsilon_2$ \\
              $0$ & $0$& $a$& $0$& $ 0$& $0$ \\
              $0$ & $0$& $-\epsilon_2 \frac{ad}{2b}$& $sa$& $0$& ${b}$ \\
              $0$ & $0$& $\epsilon_2(s\epsilon_1  b-\frac{a}{2})$& $0$& $2sb \epsilon_2 $& $d$ \\
                 \end{tabular} \right),\;\;\;\;\;\;
                 a, b\in\Re-\{0\};\;\;\;
                 c, d\in\Re,}
$$

\vspace{0.80cm}

{\bf $\cal{D}$$sd^{7\;\;p=0}_{(2,4)}:$}\\

{\footnotesize $\Big(C^5_{0} ,
(A_{1,1}+2A)^0_{0,0,\epsilon_1}\Big)\longrightarrow \Big(C^5_0 ,
(A_{1,1}+2A)^1_{k,0,\epsilon_2 }\Big),\;\;\;\;\epsilon_1,
\epsilon_2 =\pm 1 $}

$$
{\footnotesize C=\left( \begin{tabular}{cccccc}
              $1$ & $c$& $ 0$& $ 0$ & $ 0$& $ 0$\\
              $0$ & $\frac{a^2}{\epsilon_1}(k+\epsilon_2)$& $ 0$& $ 0$& $ 0$& $ 0$ \\
              $0$ & $0$& $0$& $a$& $0$& $ 0$ \\
              $0$ & $0$& $-a$& $0$& $ 0$& $ 0$ \\
              $0$ & $0$& $-\epsilon_2\frac{a}{2}$& $b$& $0$& $\frac{a}{\epsilon_1}(k+\epsilon_2)$ \\
              $0$ & $0$& $-b$& $\epsilon_2 \frac{a}{2}$& $-\frac{a}{\epsilon_1}(k+\epsilon_2)$& $0$ \\
                 \end{tabular} \right),\;\;\;\;\;\;
                 a\in\Re-\{0\};\;\;\;
                 b, c\in\Re,}
$$

\vspace{0.60cm}

{\footnotesize $\Big(C^5_{0} ,
(A_{1,1}+2A)^0_{0,0,\epsilon_1}\Big)\longrightarrow \Big(C^5_0 ,
(A_{1,1}+2A)^2_{s,0,\epsilon_2 }\Big),\;\;\;\;\epsilon_1,
\epsilon_2 =\pm 1 $}

$$
{\footnotesize C=\left( \begin{tabular}{cccccc}
              $1$ & $c$& $ 0$& $ 0$ & $ 0$& $ 0$\\
              $0$ & $\frac{a^2}{\epsilon_1}(s+\epsilon_2)$& $ 0$& $ 0$& $ 0$& $ 0$ \\
              $0$ & $0$& $0$& $a$& $0$& $ 0$ \\
              $0$ & $0$& $-a$& $0$& $ 0$& $ 0$ \\
              $0$ & $0$& $-\epsilon_2\frac{a}{2}$& $b$& $0$& $\frac{a}{\epsilon_1}(s+\epsilon_2)$ \\
              $0$ & $0$& $-b$& $\epsilon_2 \frac{a}{2}$& $-\frac{a}{\epsilon_1}(s+\epsilon_2)$& $0$ \\
                 \end{tabular} \right),\;\;\;\;\;\;
                 a\in\Re-\{0\};\;\;\;
                 b, c\in\Re,}
$$

\vspace{0.90cm}

{\bf $\cal{D}$$sd^{7\;\;p}_{(2,4)}:\;\;\;\;p>0$}\\

{\footnotesize $\Big(C^5_{p} ,
(A_{1,1}+2A)^0_{0,0,\epsilon_1}\Big)\longrightarrow \Big(C^5_p ,
(A_{1,1}+2A)^1_{k,0,\epsilon_2 }\Big),\;\;\;\;\epsilon_1,
\epsilon_2 =\pm 1 $}

$$
{\footnotesize C=\left( \begin{tabular}{cccccc}
              $1$ & $c$& $ 0$& $ 0$ & $ 0$& $ 0$\\
              $0$ & $\frac{a^2(k-\epsilon_2)-4ab(1+p^2)}{\epsilon_1}$& $ 0$& $ 0$& $ 0$& $ 0$ \\
              $0$ & $0$& $0$& $-a$& $0$& $ 0$ \\
              $0$ & $0$& $a$& $0$& $ 0$& $ 0$ \\
              $0$ & $0$& $-b$& $\frac{a\epsilon_2+2b(1+2p^2)}{2p}$& $0$& $-\frac{a(k-\epsilon_2)-4b(1+p^2)}{\epsilon_1}$ \\
              $0$ & $0$& $-\frac{a \epsilon_2+2b}{2p}$& $b$& $\frac{a(k-\epsilon_2)-4b(1+p^2)}{\epsilon_1}$& $0$ \\
                 \end{tabular} \right),\;\;
                 a\in\Re-\{0\};\;
                 b, c\in\Re,}
$$

\vspace{0.70cm}

{\footnotesize $\Big(C^5_{p} ,
(A_{1,1}+2A)^0_{0,0,\epsilon_1}\Big)\longrightarrow \Big(C^5_p ,
(A_{1,1}+2A)^2_{s,0,\epsilon_2 }\Big),\;\;\;\;\epsilon_1,
\epsilon_2 =\pm 1 $}

$$
{\footnotesize C=\left( \begin{tabular}{cccccc}
              $1$ & $c$& $ 0$& $ 0$ & $ 0$& $ 0$\\
              $0$ & $\frac{a^2(s-\epsilon_2)-4ab(1+p^2)}{\epsilon_1}$& $ 0$& $ 0$& $ 0$& $ 0$ \\
              $0$ & $0$& $0$& $-a$& $0$& $ 0$ \\
              $0$ & $0$& $a$& $0$& $ 0$& $ 0$ \\
              $0$ & $0$& $-b$& $\frac{a\epsilon_2+2b(1+2p^2)}{2p}$& $0$& $-\frac{a(s-\epsilon_2)-4b(1+p^2)}{\epsilon_1}$ \\
              $0$ & $0$& $-\frac{a \epsilon_2+2b}{2p}$& $b$& $\frac{a(s-\epsilon_2)-4b(1+p^2)}{\epsilon_1}$& $0$ \\
                 \end{tabular} \right),\;\;
                 a\in\Re-\{0\};\;
                 b, c\in\Re,}
$$

\vspace{0.70cm}

{\footnotesize $\Big(C^5_{p} ,
(A_{1,1}+2A)^0_{0,0,\epsilon}\Big)\longrightarrow \Big(C^5_p ,
I_{(1,2)}\Big),\;\;\;\;\; \epsilon =\pm 1 $}

$$
{\footnotesize C=\left( \begin{tabular}{cccccc}
              $-1$ & $c$& $ 0$& $ 0$ & $ 0$& $ 0$\\
              $0$ & $\frac{4abp(1+p^2)}{\epsilon}$& $ 0$& $ 0$& $ 0$& $ 0$ \\
              $0$ & $0$& $-a$& $2ap$& $-\frac{4ap}{\epsilon}$& $\frac{4ap^2}{\epsilon}$ \\
              $0$ & $0$& $0$& $a$& $ \frac{4ap^2}{\epsilon}$& $\frac{4ap}{\epsilon}$ \\
              $0$ & $0$& $-b$& $pb$& $0$& $0$ \\
              $0$ & $0$& $pb$& $b$& $0$& $0$ \\
                 \end{tabular} \right),\;\;
                 a, b\in\Re-\{0\};\;
                 c\in\Re,}
$$

\vspace{0.90cm}

{\bf $\cal{D}$$sd^{8}_{(2,4)}:$}\\

{\footnotesize $\Big(C^5_{0} ,
(A_{1,1}+2A)^2_{\epsilon,0,-\epsilon}\Big)\longrightarrow
\Big(C^5_0 , I_{(1,2)}\Big),\;\;\;\;\; \epsilon =\pm 1 $}

$$
{\footnotesize C=\left( \begin{tabular}{cccccc}
              $1$ & $e$& $ 0$& $ 0$ & $ 0$& $ 0$\\
              $0$ & $ab-cd$& $ 0$& $ 0$& $ 0$& $ 0$ \\
              $0$ & $0$& $\frac{\epsilon d}{2}$& $-a$& $0$& $-d$ \\
              $0$ & $0$& $a$& $-\frac{\epsilon d}{2}$& $d$& $0$ \\
              $0$ & $0$& $\frac{\epsilon b}{2}$& $-c$& $0$& $-b$ \\
              $0$ & $0$& $c$& $-\frac{\epsilon b}{2}$& $b$& $0$ \\
                 \end{tabular} \right),\;\;\;\;ab-cd \neq 0;\;\;\;
                 a, b, c, d, e\in\Re.}
$$

%%%%%%%%%%%%%%%%%%%%%%%%%%%%%%%%%%%%%%%%%%%%%%%%%%%%%%%%%%%%%%%%%%%%%%%%%%%%%%%%%%%%%%


\begin{thebibliography}{99}

\bibitem{N.A} {\footnotesize N. Andruskiewitsch, \textit {Lie Superbialgebras and Poisson-Lie supergroups},
 Abh. Math. Sem. Univ. Hamburg {\bf 63} (1993) 147-163. }

\bibitem{Bs} {\footnotesize  N. Beisert and E. Spill, \textit {The
classical r-matrices of AdS/CFT and its Lie bialgebras structure},
\texttt{[arXiv:0708.1762 [hep-th]]}.}



\bibitem {ER} {\footnotesize A. Eghbali, A. Rezaei-Aghdam, \textit
{Poisson-Lie T-dual sigma models on supermanifolds}, JHEP 09
(2009) 094, \texttt{[arXiv:0901.1592v3 [hep-th]]}. }


\bibitem {Geer} {\footnotesize N. Geer, \textit{Etingof-Kazdan quantization of Lie
super bialgebras}, \texttt{[arXiv:math.QA/0409563]}. }

\bibitem {J.z} {\footnotesize C. Juszczak and J. T. Sobczyk,  \textit {Classification of low dimentional Lie
super-bialgebras, } \textit{J. Math. Phys.} \textbf{39}, (1998)
4982-4992. }


\bibitem{J} {\footnotesize  C. Juszczak,  \textit {Classification of $osp(2|2)$ Lie super-bialgebras, }
\texttt{[arXiv:math.QA/9906101]}. }


\bibitem{KARAALI} {\footnotesize G. Karaali, \textit{ A New Lie Bialgebra Structure on
sl(2,1)}, \textit{ Contemp. Math.} \textbf{413}  (2006),
pp.101-122.}


\bibitem {ER1} {\footnotesize A. Eghbali, A. Rezaei-Aghdam and  F. Heidarpour, \textit{
Classification of two and three dimensional Lie super-bialgebras
}, J. Math. Phys. {\bf 51}, 073503 (2010);
\texttt{[arXiv:0901.4471v [math-ph]]}. }


\bibitem{B} {\footnotesize N. Backhouse, \textit{ A classification of four-dimensional Lie
superalgebras}, \textit{J. Math. Phys.} \textbf{ 19} (1978)
2400-2402.}

\bibitem {H} {\footnotesize L. Hlavaty, J. Vysoky, \textit{ Manin supertriples and
Drinfel'd superdoubles in low dimensional},
\texttt{[arXiv:0908.0997v2 [math-ph]]}. }

\bibitem {R1} {\footnotesize Note that the bracket of one boson with one boson or one fermion
is usual commutator but for one fermion with one fermion is
anticommutator. Furthermore we identify grading of indices by the
same indices in the power of (-1), for example $grading(i)\equiv
i$; this is the notation that DeWitt applied in his book (Ref.
\cite{D}).}


\bibitem{D} {\footnotesize B. DeWitt, \textit{ Supermanifolds,} (Cambridge University Press
1992).}

\bibitem{RHR}{\footnotesize  A. Rezaei-Aghdam, M. Hemmati, A. R. Rastkar, \textit{
classification of real three-dimensional Lie bialgebras and their
Poisson-Lie groups}, \textit{ J. Phys. A:Math.Gen.} \textbf{ 38}
(2005) 3981-3994, \texttt{[arXiv:math-ph/0412092]}. }

\bibitem{JR} {\footnotesize M.A. Jafarizadeh and A. Rezaei-Aghdam, \textit{ Poisson-Lie
T-duality and Bianchi type algebras}, \textit{Phys. Lett. B}
\textbf{ 458} (1999) 477-490, \texttt{[arXiv:hep-th/9903152]}. }

\bibitem {Snobl} {\footnotesize L. Snobl, L. Hlavaty,
\textit{ Classification of 6-dimensional real Drinfeld doubles},
\textit{ Int.J.Mod.Phys.} \textbf{ A17} (2002) 4043-4068,
\texttt{[arXiv:math.QA/0202210v2]}. }

\bibitem {R2} {\footnotesize The Lie superalgebra $A$ is  one dimensional Abelian Lie
superalgebra with one fermionic generator where Lie superalgebra
$A_{1,1}$ is its bosonization. Furthermore, $C^1_{\frac{1}{2}}$
is different from $C^1_p$ for $p=\frac{1}{2}$ and we show the
latter by $C^1_{p=\frac{1}{2}}$. }

\bibitem {R3} {\footnotesize Note that 17 of them are  two bosons-one fermion and 53 of them
are one boson-two fermions Lie super-bialgebras.}

\end{thebibliography}
\end{document}